\documentclass[final,5p,times,twocolumn]{elsarticle}

\usepackage{amssymb}
\usepackage{amsmath}
\usepackage{amsmath,amsfonts,bm}

\def\eqref#1{equation~\ref{#1}}
\def\1{\bm{1}}

\def\ra{{\textnormal{a}}}

\def\rva{{\mathbf{a}}}

\def\va{{\bm{a}}}

\def\vd{{\bm{d}}}

\def\vq{{\bm{q}}}
\def\vr{{\bm{r}}}
\def\vs{{\bm{s}}}

\def\vv{{\bm{v}}}

\def\mA{{\bm{A}}}

\def\mP{{\bm{P}}}
\def\mQ{{\bm{Q}}}

\DeclareMathAlphabet{\mathsfit}{\encodingdefault}{\sfdefault}{m}{sl}
\SetMathAlphabet{\mathsfit}{bold}{\encodingdefault}{\sfdefault}{bx}{n}
\newcommand{\tens}[1]{\bm{\mathsfit{#1}}}

\def\tQ{{\tens{Q}}}

\def\sQ{{\mathbb{Q}}}

\def\sS{{\mathbb{S}}}

\newcommand{\etens}[1]{\mathsfit{#1}}

\def\etQ{{\etens{Q}}}

\newcommand{\E}{\mathbb{E}}

\newcommand{\R}{\mathbb{R}}

\newcommand{\KL}{D_{\mathrm{KL}}}

\DeclareMathOperator*{\argmax}{arg\,max}
\DeclareMathOperator*{\argmin}{arg\,min}

\usepackage{blkarray}
\usepackage{subcaption}
\usepackage{wrapfig}
\usepackage[colorlinks, citecolor=blue]{hyperref}

\usepackage{amsmath}
\usepackage{amssymb}
\usepackage{mathtools}
\usepackage{amsthm}
\usepackage{algorithm}
\usepackage{algorithmicx}
\usepackage[noend]{algpseudocode}
\usepackage{xcolor}

\biboptions{sort&compress}

\renewcommand\appendixautorefname[1]{} % fix Appendix Appendix

\theoremstyle{plain}
\newtheorem{theorem}{Theorem}[section]

\theoremstyle{definition}
\newtheorem{definition}[theorem]{Definition}
\newtheorem{assumption}[theorem]{Assumption}
\theoremstyle{remark}

\journal{Neurocomputing}

\begin{document}

\begin{frontmatter}



\title{Overcoming the Price of Anarchy by Steering with Recommendations}


\author{Cesare Carissimo}
\author{Marcin Korecki}
\author{Damian Dailisan}%% Author name

\affiliation{organization={ETHZ},%Department and Organization
            city={Zurich},
            country={Switzerland}}

%% Abstract
\begin{abstract}
Varied real world systems such as transportation networks, supply chains and energy grids present coordination problems where many agents must learn to share resources. It is well known that the independent and selfish interactions of agents in these systems may lead to inefficiencies, often referred to as the `Price of Anarchy'. Effective interventions that reduce the Price of Anarchy while preserving individual autonomy are of great interest. In this paper we explore recommender systems as one such intervention mechanism. We start with the Braess Paradox, a congestion game model of a routing problem related to traffic on roads, packets on the internet, and electricity on power grids. Following recent literature, we model the interactions of agents as a repeated game between $Q$-learners, a common type of reinforcement learning agents. This work introduces the Learning Dynamic Manipulation Problem, where an external recommender system can strategically trigger behavior by picking the states observed by $Q$-learners during learning. Our computational contribution demonstrates that appropriately chosen recommendations can robustly steer the system towards convergence to the social optimum, even for many players. Our theoretical and empirical results highlight that increases in the recommendation space can increase the steering potential of a recommender system, which should be considered in the design of recommender systems.
\end{abstract}


% \begin{highlights}
% % must be 85 characters including spaces
% \item Modeled the interplay of agents playing a repeated congestion game and a recommender system.
% \item We model agents as independent Q-learners.
% \item The recommender system provides signals to optimize system efficiency.
% % \item Q-learners play a repeated congestion game and an external recommender system tries to optimize system efficiency.
% \item Recommendations are modeled as states that are selected by the external recommender.
% % \item We demonstrate that the recommender can steer learning agents away from the Nash equilibrium towards the social optimum.
% \item Recommendations can steer agents away from the Nash equilibrium towards the social optimum.

% \end{highlights}

%% Keywords

\end{frontmatter}



\section{Introduction}
\label{introduction}

Although computationally more powerful than humans, machine intelligences are not impervious to the dilemmas of competitive strategic behavior. These dilemmas, abstracted as a `Price of Anarchy', stem from the misalignment of individual and social objectives. 

% Pollution, overproduction of waste, congestion in traffic, and tragedies of the commons can be cast as challenges that stem from misaligned individual and social objectives.

As ``[m]achines powered by artificial intelligence increasingly mediate our social, cultural, economic and political interactions[,] understanding the behavior of artificial intelligence systems is essential to our ability to control their actions, reap their benefits and minimize their harms'' \cite{rahwan2019machine}. 
% new intro context, cesare pls review

Indeed, machine learning algorithms play increasingly larger roles in mediating complex, social, and human interactions.
In systems with a high Price of Anarchy, centralizing the actions of individuals to steer a system to achieve socially optimal objectives raises questions on the tradeoffs between the potential risks and benefits.
To what extent should algorithms dictate beyond content to behavior?
In this paper, we present a hypothetical scenario as a thought experiment to study this complicated real-world scenario.

% %As such, we cast this thought experiment in greater detail next.

\subsection{The Narrative Thought Experiment}

In the world of this thought experiment, it is common for intelligent machines to assist people. In many cases, decisions are delegated to machines entirely. `Under the hood', these intelligent machines are methodologically similar to the machine learning technologies available today (2025). Their widespread success and implementation arise centrally from the striking efficiency that results from their ubiquitous implementation.

In this world, the transportation sector has experienced a drastic upheaval. All vehicles implement reinforcement learning algorithms to determine their paths as they shuffle around people and resources. Though it is known that selfish decision making in complex road networks can lead to inefficient outcomes, the unprecedented speed with which automated vehicles took over road networks left little time for more holistic and coordinated implementations, so all vehicles act independently to minimize their own travel times. As it results, the system is not, on a global scale, optimally efficient: it suffers a so called Price of Anarchy. In fact, all vehicles would be better off in the long-term if they were better coordinated with each other for a small myopic sacrifice.

A new algorithm is designed to attempt a solution: a recommender system which can provide all the vehicles with additional information while the vehicles learn to route through cities. This recommender system is unprecedentedly powerful and is capable of modeling the decision making of the vehicles to determine what they would do conditional on their recommendation. To what extent can this recommender system steer the decision making of individual vehicles and overcome their Price of Anarchy? This paper explores such a recommender system, its opportunities, and its limitations.

\subsection{Overview}
\begin{figure}[!tbh]
    \centering
    \includegraphics[width=0.9\linewidth]{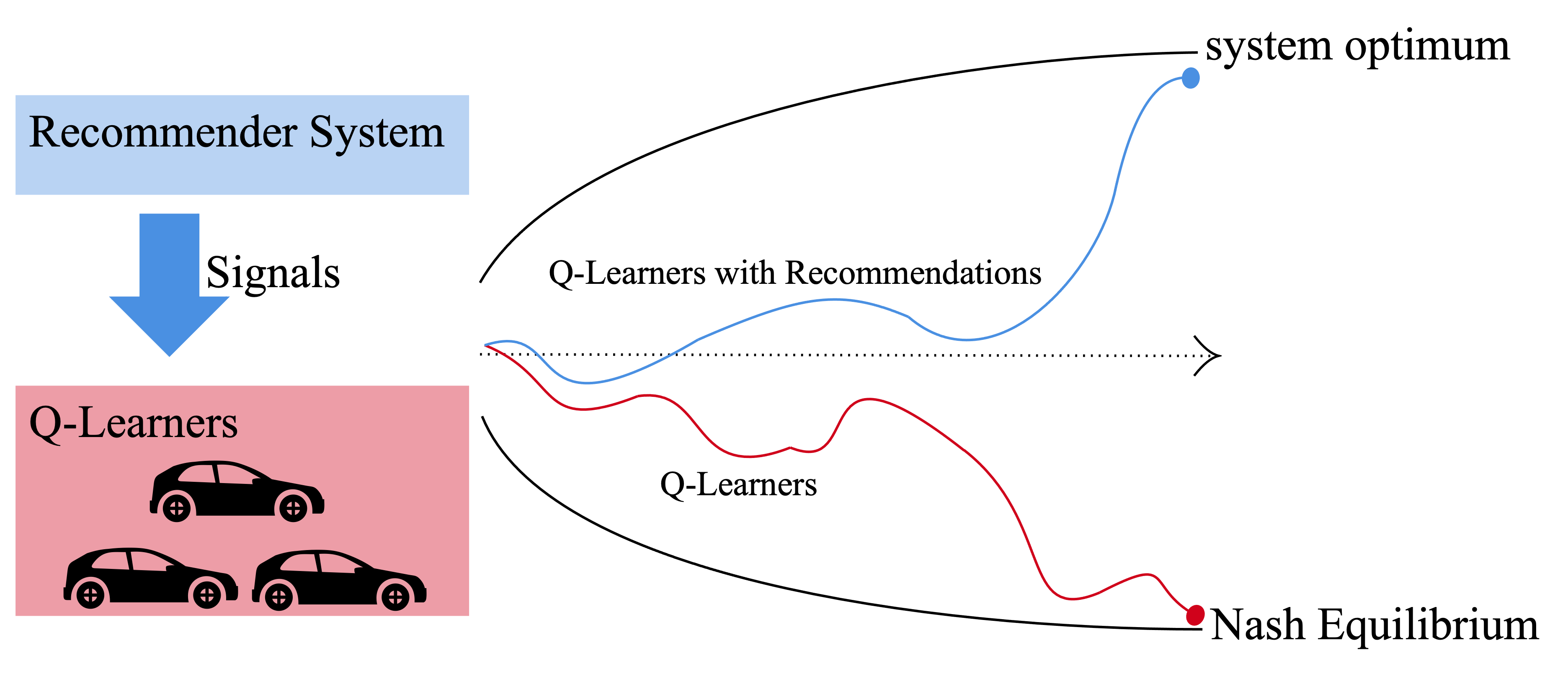}
    \caption{Graphical depiction of the main innovation and contribution: steering a multi-agent system of $Q$-learners with a recommender system towards a system optimum.}
    \label{fig:ldmp_diagram}
\end{figure}

Our paper focuses on studying the interaction of a system of learning agents with an overarching system that sends them signals, or recommendations.
In this work, we use game theory and machine learning algorithms to study the extent to which we can mitigate the negative social externalities of self-interested learning agents. 
Specifically, we study a congestion game with a high Price of Anarchy. 
% This allows us to accurately model the effects of individual decision making on the system. The solution concept we discuss is that of recommendations.
%Our proposed model does not directly reflect an existing real-world system, but rather models a hypothetical scenario as a thought experiment. 
The agents are modeled as $Q$-learners. The recommender interacts with the $Q$-learners by setting their states during learning. Then we devise a deterministic algorithm to steer the system towards the social optimum, as depicted in \autoref{fig:ldmp_diagram}. 

We choose the Braess Paradox, a congestion game with a high Price of Anarchy, where uncoordinated agents have a strong tendency to end up in a Pareto inefficient Nash equilibrium.
For the computational results (\autoref{sec:results2}), we generalize the recommendations to abstract `signals' which can be more than the actions available to $Q$-learners.
We show that our recommender system can robustly steer the learning dynamics with strategic signals, even for many agents, without directly controlling the choices of agents.
Although we study a hypothetical scenario, our model may provide insights for practical implementations of routing recommendations where the agents are learning and may not necessarily follow the recommendations.

% We now proceed as follows: In \autoref{sec:theory} we discuss our model. In \autoref{sec:results1} we look at recommendations given to $Q$-learners for routes ('paths') in Braess' Paradox. We show the possibility to enhance performance through recommendations and discuss the limitations of this approach related to noise and alignment. In \autoref{sec:results2} we generalize the notion of recommendations to signals that are no longer strictly tied to available routes, but are treated as meaningful information for learning. By this, we demonstrate enhanced possibilities to manipulate the learning dynamics through strategic signals. In \autoref{sec:discussion} we discuss the results and implications of this thought experiment.

Summary of contributions:
\begin{itemize}
    \item A model of a plausible scenario where a recommender systems seeks to control the behaviour of agents in a congestion game.
    \item Assuming that agents have dynamically changing behaviour by modeling them as Q-learners.
    \item Creating an algorithm that can steer the learning dynamics of a system of many agents, which can drive the system towards the social optimum, while being robust to the number of agents.
\end{itemize}

\section{Related Work}
\label{sec:relatedwork}
In this section we discuss the theoretical concepts and literature that are relevant to a recommender system which interacts with reinforcement learners that play a repeated congestion game with a large Price of Anarchy.

\subsection{Modeling Congestion}

Traffic is a ubiquitous complex system driven by the interactions of vehicles on road networks. Congestion is the particular state of traffic where the road network is `saturated' in that it reaches its maximum capacity, the limit of vehicles that can simultaneously use the roads. When traffic is not congested, the individual choices of vehicles are less influential to traffic flows. Conversely, when traffic is congested the influence of individual vehicle choices is heightened. 

It is well understood that the complex physical interactions between vehicles can create worsen congestion \cite{bando1995dynamical, helbing2002micro}. For example, traffic phenomena like phantom jams and green waves could be explained by these microscopic interactions between vehicles \cite{helbing2001traffic}. Vehicles are modeled as pursuing a target speed and slowing down when they approach a neighboring vehicle, but with a reaction delay. 

In this paper, we focus on the effects of route choice on congestion. Which paths do vehicles pick in a traffic network \cite{wardrop1952road}? We start by assuming that each vehicle prefers a shorter travel time. Then we conclude that each vehicle will select the shortest path available to it. In so doing we enter the dynamic realm of game theory where individual path choices may influence all travel times; reasonably, if all vehicles pick the same shortest path it may stop being the fastest. This gives rise to the Traffic Assignment Problem which requires picking a path between origins and destinations for the vehicles in a road network. Wardrop's seminal work on the subject \cite{wardrop1952road} formalized a traffic assignment reached by selfish vehicles as the User Equilibrium, and one coordinated by a central planner as the Wardrop Equilibrium.

% While the dynamics of traffic congestion driven by route choices are known to depend largely on the road network topology. Despite this, most roads in cities were not designed with widespread GPS routing recommendations and their effects in mind. It is known that congestion may worsen, when all users selfishly choose the same shortest paths.

\subsection{Congestion Games}

Routing, route choice, and congestion are a prominent part of game theory \cite{rosenthal1973class}. These `congestion games' assume that $N$ players share a set $A$ of roads where selecting a resource $a$ yields a utility $u_a(f(a))$, which depends on the number of players $f(a)$ that selected road $a$ and $u_a$ is a non-increasing function (more players on the same road lead to less utility per player). When we assume that players are rational utility maximizers (travel time minimizers), game theory defines solution concepts like Nash equilibria to characterize the road network (equivalent to the Nash Equilibrium (NE) for non-atomic players \cite{wie1998relationship}). Congestion games are a special case of convex potential games \cite{monderer1996potential} --- games that admit a convex potential function whose maximization leads to the NE --- and as such have a unique pure strategy NE. The unique pure NE tells us that if each player picks the NE path, no individual player is incentivized to pick a different path. 

As formulated, the NE may not be the best description of the state of a real traffic system, as it may be reasonable to assume that the vehicles are unaware of the NE strategy, or that they have partial information about what other vehicles are doing. Perhaps it is reasonable to assume that each vehicle receives some additional information on which they can decide what to do, such as a route recommendation. This situation has a more suitable economic solution concept, known as a correlated equilibrium. A correlated equilibrium is defined as an NE with additional information given to all players, just like recommendations. In a correlated equilibrium, no player is incentivized to change their action given their recommendation and given that they know what all other players are recommended. Strikingly, for non-atomic congestion games, it is proven that correlated equilibria correspond exactly with the Nash Equilibria \cite{koessler2024correlated}. If this result were a general fact of route choice in real-world traffic systems it would entail that: route recommendations can not change the route choice of vehicles in a road network. Such a prediction is hardly justified. Route recommendations from modern platforms like Google Maps and Waze clearly influence vehicles in traffic. The kind of player that is unaffected by a route recommendation is one who ignores it. 

A key feature of this work is the introduction of recommendations as an input to route choices.
% In this paper, this will be a central feature of the route choices in our model: they will be affected by recommendations. 
Furthermore, agents will repeatedly play a congestion game and learn to do so over time. In repeated congestion games, selfish behavior can sustain outcomes that are better than NE in terms of social welfare (Pareto-optimal) \cite{scarsini2012repeated}. At the same time, Best-Response dynamics are known to converge in congestion games \cite{candogan2013near}. It is relevant, therefore, to look beyond Best-Response dynamics if we wish to study behavior away from the equilibrium. 
% Correlated equilibria are a well-known generalization of NE \cite{aumann1987correlated}, which contain equilibria that are possibly better than NE. A correlated equilibrium is achieved with a correlation device, meaning that all players are given additional information, which can change their best responses. A correlation device can be understood as a RS. For rational players and Best-Response dynamics in convex potential games, there are expected to be little possible improvements with correlation devices \cite{roughgarden2015intrinsic, koessler2023correlated}.

\subsection{Braess's Paradox}

\begin{figure*}[!ht]
\centering
\begin{subfigure}{.4\textwidth}
  \centering
  \includegraphics[width=0.8\linewidth]{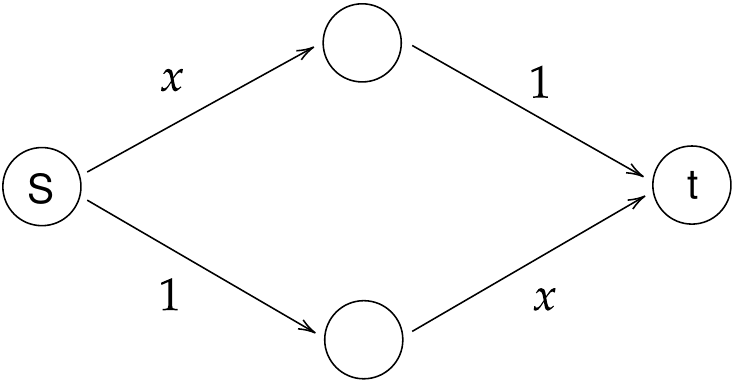}
  \caption{Initial Network}
  \label{fig:sub1}
\end{subfigure}%
\begin{subfigure}{.4\textwidth}
  \centering
  \includegraphics[width=0.8\linewidth]{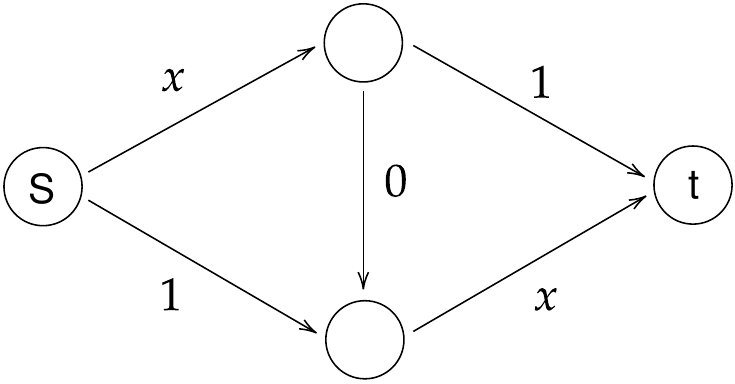}
  \caption{Augmented Network}
  \label{fig:sub2}
\end{subfigure}
\caption{Illustration of the initial network (a), and the augmented network (b) in the Braess Paradox. Agents start in the ``S'' state and pick a path to reach state ``t''. The numbers represent the cost of traveling over a link. A cost of $x$ is the ratio of agents that choose that link. Two actions are possible in (a), $\mathit{up}$ takes the upper edges, and $\mathit{down}$ takes the lower edges. In (b) an additional action $\mathit{cross}$ is possible, which takes the first upper edge, crosses to the lower section at the middle, and finishes on the second lower edge. Rational and fully-informed agents all pick the crossing link in the augmented network (Nash equilibrium), which leads to high congestion and the worst possible social welfare.}
\label{fig:braess_network}
\end{figure*}

In practice, vehicles do not use roads in the most efficient way. In fact, most concern stems from the realization that self-interested choices can worsen congestion. In economic terms there are road networks which have a Pareto inefficient NE. The Pareto inefficiency entails that there exists an assignment of traffic which improves the travel time for all vehicles without worsening the travel time for any vehicle. This inefficiency is exemplified in the Braess Paradox, whereby the capacity of a road network can be increased while simultaneously worsening the NE social welfare, and subsequently creating a network where self-interested vehicles make congestion worse than when there was less road capacity. This counter-intuitive phenomenon is also found to have real-life relevance for the routing of packets on the Internet~\cite{korilis1999avoiding, tumer2000collective}, the flow of energy on power grids \cite{schafer2022understanding}, and the path choice of drivers in urban road networks \cite{argota4291171less}. In this paper we will be using the Braess Network for our investigations as captures in \autoref{fig:braess_network}.

% \begin{figure}
%     \centering
%     \includegraphics[width=0.6\linewidth]{figures/braess_with_eq.pdf}
%     \caption{Illustration of the Braess Paradox network. Agents start in the ``s'' state and pick a path to reach state ``t''. The numbers represent the cost of traveling over a link. A cost of $x$ is the ratio of agents that choose that link. Rational and fully-informed agents all pick the crossing link in the augmented network (NE), which leads to high congestion and the worst possible social welfare.}
%     \label{fig:braess_network}
% \end{figure}

The Braess Paradox has a structure similar to other popular games in game theory, like the Prisoner's Dilemma, the Public Goods game and the Bertrand Duopoly. These games share the property that the Nash Equilibrium is inefficient, meaning that there are better collusive that the players could establish if they could convince each other not to be selfish, or alternatively if they were forced not to be.

\subsection{Price of Anarchy}

This aforementioned property of an inefficient Nash Equilibrium which many games share is also called the Price of Anarchy \cite{koutsoupias1999worst}, and was first explored for congestion games. First, a suitable system wide metric is defined, like social welfare, which is taken to be the average (or sum) of all the utilities of the players. Then, the Price of Anarchy is defined as the ratio of the social welfare at the worst Nash Equilibrium (NE) and the socially optimal solution. Intuitively, it is referred to as the cost of having self-interested decision makers rather than coordinated social control. If systems like traffic road networks did not have a Price of Anarchy, it may the theoretically justifiable not to intervene at all. However, given the widespread identification of theoretical networks with a Price of Anarchy larger than 1, there is a theoretical `room for improvement'. The question is how much. For linear congestion games the Price of Anarchy is upper bounded to $\frac{4}{3}$, where the augmented network of the Braess Paradox is the network which saturates this upper bound. In other words, in the worst case road network self-interested route choice is $33\%$ worse than the social optimum.

Now, depending on the field of interest a worst case inefficiency of $33\%$ may have different interpretations. How bad is this, really? Within the congestion game literature, this inefficiency is interpreted with varying severity. The authors of these bounds viewed them optimistically as proof of a limit to the inefficiency of self-interested behavior, at least for congestion games \cite{roughgarden2002bad}. Furthermore, if the assumption of linear costs seems unreasonable (which it certainly will to traffic scientists which more commonly use a BPR cost-function \cite{us1964traffic}) the same paper established an upper bound of $2$ for the Price of Anarchy when the costs functions were only restricted to be increasing in the number of players, once again in the very worst case.

If this is `as bad as it gets', then what features of vehicle behavior may reduce the Price of Anarchy? After all, the assumptions of rational behavior required to derive these worst-case bounds are easily criticized for their stringency. Recent work has challenged the Price of Anarchy in the Braess Paradox by deploying reasonable reinforcement learning algorithms in the road network and demonstrating that even their learning dynamics --- though self-interested --- lead to a significantly better social welfare than rational agents \cite{carissimo2024counter}. This paper constitutes an extension of that work, the discussion of which we defer to the next sections.

The upper bounds on the Price of Anarchy remain to be interpreted. In this paper, as we seek to mitigate the Price of Anarchy with a coordination algorithm, we will frequently refer to the mirrored perspective which can be though of as a `Cost of Control'. How costly --- to solve, to implement, in its externalities --- is the coordination solution required to mitigate the inefficiencies of self-interested behavior? It is often the case that, while possibly imperfect, coordination solutions via self-organization mechanisms are most efficient.

\subsection{Road Pricing}

A common approach to improving the efficiency of traffic and reducing congestion is to charge drivers for their road choices \cite{morrison1986survey}. By making roads more expensive it is believed that drivers choices can be shifted to the more efficient roads. Marginal cost pricing is an economically effective way of achieving these results. They rest on assumptions that drivers will be impacted by these costs. Algorithmic drivers which act in approximately rational ways may well conform to these models. Comprehensive reviews of road pricing literature find the benefits to exceed the costs \cite{anas2011reducing}. An oft raised issue with road pricing is the potential un-fairness of their results when applied to systems with unequal initial distributions of capital, where few agents may be less affected by an increase in price that others \cite{levinson2010equity}. Furthermore, the evaluation of road pricing is conducted empirically but is primarily due to reductions in overall demand, rather than the re-allocation of resource use to more efficient paths. In fact, the success of road pricing is greatly influenced by the availability of public transport options \cite{anas2011reducing}. As such, the success of road pricing is less due to a reduction of the Price of Anarchy, as it is to the reduction of demand. It is less clear and a challenge to evaluate whether road pricing is an effective method to reduce the Price of Anarchy.

% \subsection{Cybernetic Societies}

Is there a way to manage road networks in congested states without the detour of money? Road pricing schemes feature money as the social coordination tool that can re-distribute vehicles away from inefficient path choices. As such, road pricing acts on the 
incentives of agents. Would it be enough to provide agents with additional information? 
% This information should be calibrated just right to produce a desired action. From this approach we can ask whether there exists an individualized distribution of information such that the vehicles pick different paths. Cybernetics recognizes complex dynamics and feedback loops present in systems that govern their behavior, and asks whether one can intervene to steer a system. If vehicles react to perceptual inputs, which inputs lead to the best social welfare outcomes? 

% From such a perspective, each vehicle is represented as a function, a mapping from perceptions to actions. If this map is known it is possible to determine \textit{a priori} which perceptions lead to a desired path choice. If this function is not known, it may be learned by interaction with the vehicle. Over time, an observer may notice all the ways in which a vehicle responds to inputs, and learn the vehicles mapping of inputs to paths. Then, if the observer can influence the perceptual inputs of the vehicle it can influence the paths that the vehicle will choose. If the observer can do so for all vehicles it can influence the social welfare of the traffic system. 

In this paper we investigate such an approach and method when applied to a system of learning algorithms. To this end, we will use reinforcement learning as the constrained choice model employed by the vehicles.

\subsection{Reinforcement Learning Agents}
\label{sec:stateless}
A reinforcement learning agent is an agent which learns a reward function while interacting with an environment. The environment is characterized as a state space $\mathcal{S}$ and an action space $\mathcal{A}$ which represent everything that can happen. The reward function is feedback that a learner gets for performing an action in a given state and ending up in a new state. Then a reinforcement learner, in particular a model-free reinforcement learner, can try to maximize the rewards it receives from the environment using a behavioral policy which explores its available actions and exploits those actions which provide the highest rewards \cite{sutton2018reinforcement}.

% Reinforcement learning has led to breakthrough performances like algorithms which beat humans at the games of Go and Starcraft. 
In this paper we will use $Q$-learning, a reinforcement learning algorithm useful for its simplicity and broad application potential. $Q$-learning is known to converge to `optimal behavior' in finite Markovian environments (Markov Decision Processes) when all states are visited infinitely many times \cite{watkins1992q}. Optimal behavior in this case means that the agent learns how to maximize the rewards it accumulates by interacting with the environment. In multi-agent settings, however, there are no convergence guarantees for $Q$-learning.

As we will be interested to use $Q$-learning in a multi-agent game, the lack of convergence guarantees may at first seem to be a hindrance. However, this is precisely the opposite, because agents which converge may have the same behavior regardless of the inputs they receive, just like rational self-interested agents which converge to Nash Equilibria in non-atomic congestion games no matter what additional information they are given. We will explore this feature further in this paper, but for now it suffices to claim that convergence is not the only interesting behavior that arises from the interactions of agents. Here, as we wish to study Machine Behavior, we interest ourselves in those dynamics which lead us away from equilibrium, as was previously explored in the Braess Paradox in \cite{carissimo2024counter}.

Our methodological perspective has some similarities with other literature on steering in Reinforcement Learning. Steering refers to influencing agents during their learning to guide their convergence to desired policies. For example, single agent steering with multiple criteria \cite{mannor2001steering}, steering Markovian agents \cite{huang2024learning}, and  steering no-regret learners \cite{zhang2023steering}. Steering typically involves adjusting the reward function during learning. Our approach differs in that we adjust the state information that reinforcement learners observe during learning.

We note that no-regret learning is a popular model for dynamic agents which interact with an environment with appealing properties \cite{hart2000simple}. A no-regret learner is defined as an agent which, in the limit of infinite time, will converge to playing the actions which are optimal. This agent may explore during the course of its learning, in the beginning for example, but is assumed to eventually play the optimal actions. We do not assume no-regret learners, or analyze our $Q$-learners against regret. This is because in our game setting, a no-regret learner converges to the Nash Equilibrium behavior. Furthermore, we hypothesize that a no-regret can not be steered with recommendations, even though they can be steered with rewards \cite{zhang2023steering}. Therefore, to study steering with recommendations we look beyond no-regret learning and settle on $Q$-learning.

% \subsection{Congestion Games with Learning}\label{sec:stateless}
This paper analyses a repeated congestion game with $Q$-learning agents \footnote{The $Q$-learning agents can also be seen as weakly rational agents, as agents with bounded rationality}. The congestion game can be framed as a Markov Decision Process (MDP), with the actions corresponding to the set of resources $\mathcal{A}$ and the reward function of the MDP being equal to the utilities experienced by the agents $R(\va) = \left(u_{a_1}\left(f(a_1)\right), \dots, u_{a_n}\left(f(a_n)\right)\right)$. The $Q$-learners are assumed to have an $\epsilon$-greedy policy, $\argmax$ with probability $1-\epsilon$ and uniform random with probability $\epsilon$, and update their $q$-values with the Bellman update rule. The $Q$-learners have their own state--action value function $Q_{i}: \mathcal{S} \times \mathcal{A} \rightarrow \R$, a policy function $\pi_i: \mathcal{S} \rightarrow \mathcal{A}$, and an update rule $U_i: \mathcal{S} \times \mathcal{A} \times \R \times \mathcal{S} \rightarrow \R$ parametrized with learning rate $\alpha$ and discount factor $\gamma$. The environment runs for $\tau$ steps, and at each step $t$ $Q$-learners apply a policy $\pi_i$ to determine their next action $a_{i,t}$, observe $(s_{i,t}, a_{i,t}, r_{i,t}, s'_{i,t})$ and then update $Q_{i,t}$ using their update rule $U_i$ to obtain $Q_{i,t+1}$.

\begin{figure}[!ht]
\centering
\includegraphics[width=\linewidth]{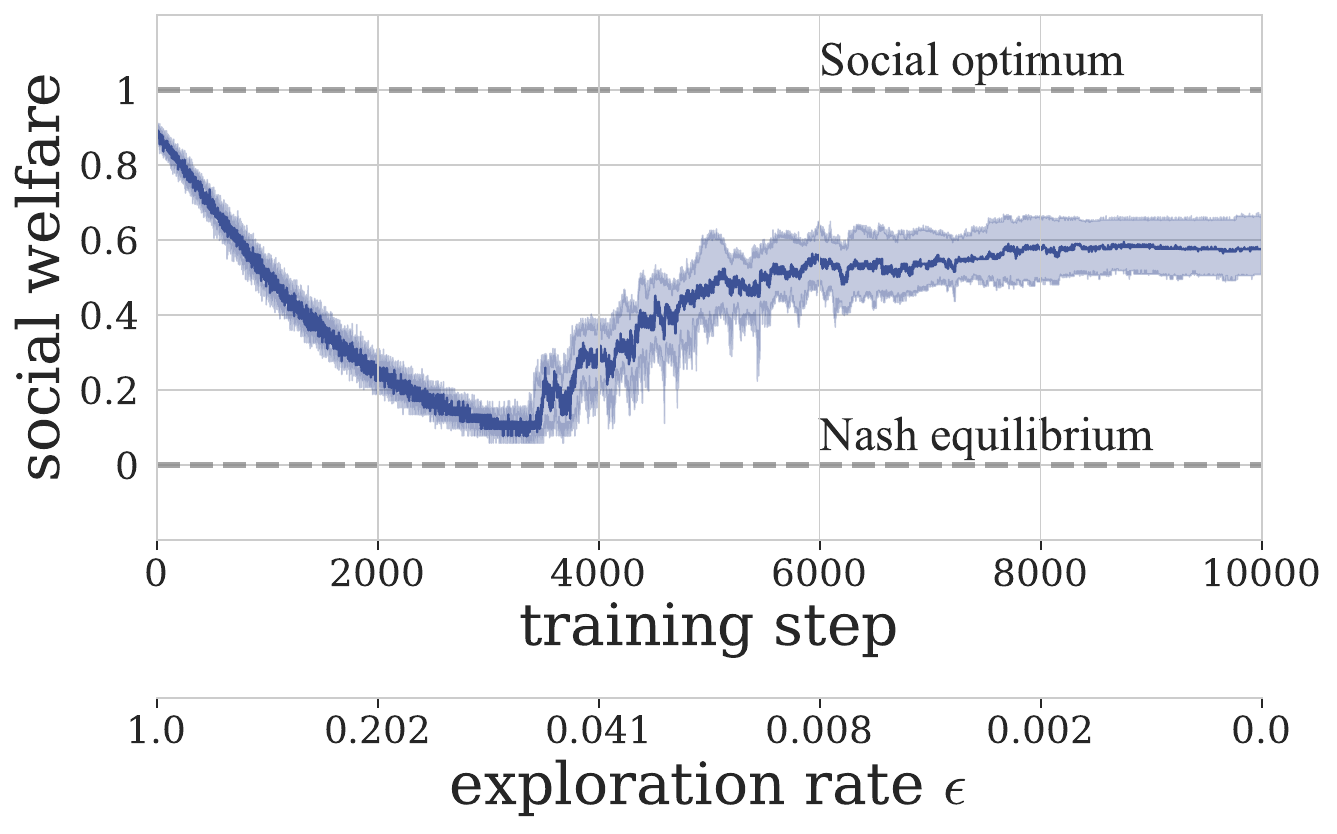} \\
\caption{Learning of 100 $\epsilon$-greedy tabular $Q$-learners \mbox{($\alpha=0.1$, $\gamma=0.8$)} in the Braess Paradox converges to social welfare values much higher than the NE (0). Values of social welfare were rescaled from \mbox{$[-2,-1.5]\rightarrow[0,1]$}, higher social welfare is better. Results replicated from \cite{carissimo2024counter}.}
\label{fig:braess_network_learning}
\end{figure}

Independent reinforcement learners that apply incremental updates to their policies are drawn towards the NE, but the NE may not be a stable equilibrium \cite{kleinberg2011beyond, bielawski2021follow, bielawski2022route}. Additionally, the NE may not be the optimum of the social welfare function defined as $W=\frac{1}{n}\sum_{i}^n r_i$. It has already been shown, for some games (pricing game \cite{calvano2020artificial, klein2021autonomous}, prisoner's dilemma \cite{schaefer2022emergence, dolgopolov2022reinforcement}), that $Q$-learners are able to perform better social outcomes than the NE through implicit coordination. These same effects have been shown in the Braess Paradox \cite{carissimo2024counter}, which we replicate in \autoref{fig:braess_network_learning}. It was demonstrated that $Q$-learning in the Braess Paradox could be chaotic and oscillatory while leading to better social welfare \cite{carissimo2024counter}. In this paper we seek to amplify these effects by steering with strategic state recommendations.

Braess's Paradox has a Nash Equilibrium where all agents cross, which leads to an average latency of $2$. However, the Social Welfare optimizing solution is for half of the agents to go up and the other half to go down, for an average latency of $1.5$. The latencies experienced by the agents are linear in the fraction of agents that choose the actions, $\frac{n_{u}}{N}, \frac{n_{d}}{N}$. Specifically, the latencies, $l(a)$, are: $l(u) = 1 + \frac{n_{u}+n_{c}}{N}, l(d) = 1 + \frac{n_{d}+n_{c}}{N}, l(c) = \frac{n_{u}+n_{c}}{N} + \frac{n_{d}+n_{c}}{N}$. The rewards for agents are then the negative of the latency ($r_i=-l(a_i)$). \autoref{fig:braess} in the appendix gives additional information about the learning dynamics in the augmented network of the Braess paradox, which are not the focus of this paper.

\subsection{Recommender Systems}

Route recommender systems are a prominent example of recommender systems (RS) that promise to reduce average travel times for drivers. In the simplest case, shortest path algorithms are applied directly to networking and traffic \cite{abolhasan2004review, zeng2009finding}, and extended to user platforms such as Google Maps and Waze \cite{luxen2011real, wang2014r3, dai2015personalized}. This approach, however, runs into trouble when all drivers receive the same shortest path recommendation and follow it: then, the recommended path will often get congested and slow drivers down. If drivers realize this, they might `learn' not to follow the recommendations \cite{helbing2002volatile}. 

Therefore, good route recommender systems (RRSs) must account for the collective effects they trigger when recommending routes. While it is widely accepted that RRSs influence the systems they interact with and that users are not static entities with fixed preferences \cite{Nguyen2014, Stocker2020, Stray2020, stray2021you, stray2021designing, Hazrati2022}, it is still not entirely clear how users should be modeled dynamically \cite{jiang2019degenerate, chen2019generative, kalimeris2021preference}. Furthermore, it is recognized that the rules of digital platforms create incentives for agents \cite{benporat2018gametheoretic, hron2023modeling}. It is an active area of research to model the effects that RRSs have on the social welfare of the systems they interact with \cite{helbing2004dynamic}. For example, by establishing metrics that can predict actually resulting performance \cite{krauth2020offline}, creating game-theoretic models where users are utility maximizers rather than static entities with fixed preferences \cite{chen2019generative}, allowing user preferences to be shaped by recommendations \cite{kalimeris2021preference}, and understanding the emergence of `echo chambers' and `filter bubbles' \cite{jiang2019degenerate}. 

Route recommendation methods are increasingly common \cite{su2014crowdplanner, cui2018personalized, su2009survey}. They have the advantage of being able to rely on traffic flow models and congestion games to estimate their validity. With game-theoretic solution concepts, researchers have shown that ubiquitous shortest path planning may cause or worsen traffic congestion \cite{thai2016negative, cabannes2017impact, macfarlane2019apps}, and possibly affect cities' economies \cite{sweet2011does}. 

Finally, the model of recommendation which we explain in the next section is novel but has some similarity to previous work on machine teaching \cite{zhu2015machine, zhu2018overview}. The goal of machine teaching is to select the optimal dataset that will allow a learner to learn the optimum. Recent work has extended machine teaching to reinforcement learning \cite{lewandowski2022reinforcement}. Our work is similar to machine teaching because we assume knowledge of a target distribution (like an optimum) to which we would like the system of $Q$-learning agents to converge on. It is also similar to machine teaching because the recommender in our model is capable of influencing the experiences of agents (picking data). Our model crucially differs from machine teaching because it involves picking the states for $Q$-learners rather than feeding them data (a state, action, reward, next state tuple for a reinforcement learner). Furthermore, our model has many $Q$-learners interacting in a congestion game such that what is optimal for an individual learner may not be socially optimal, leading to pluralistic notions of optimality.

\section{Model: Learning Dynamics Manipulation Problem in Repeated Congestion Games}\label{sec:theory}

In this section, we discuss the main contribution of this paper, namely a model of recommendations for $Q$-learners in congestion games. We assume that $Q$-learners are playing repeated congestion games and that a route recommender system (RS) steers the $Q$-learners with recommendations. $Q$-learners receive recommendations as their states.

In \autoref{sec:rl} we formulate steering of $Q$-learners, the Learning Dynamic Manipulation Problem (LDMP), as finding the optimal policy of a Markov Decision Process. In \autoref{sec:states} we discuss our main assumption for the modeling of recommendations. In \autoref{sec:scaling} we discuss how the LDMP scales as players ($n$), actions ($k$) or recommendations ($m$) grow. In \autoref{sec:proofs} we provide arguments to justify the design principles of the heuristic algorithm we designed to steer the $Q$-learners.

\begin{table}[ht]
    \centering
    \begin{tabular}{c|c}
        symbol & meaning \\
        \hline
        $n$ & number of agents \\
        $k$ & number of actions \\
        $m$ & number of recommendation states \\
        $\sS$ & recommendation set \\
        $s$ & recommendation \\
        $\mathcal{A}$ & action set \\
        $a$ & action \\
        $\va$ & action profile (vector) \\
        $\tQ$ & $q$-tables of all agents \\
        % $\etQ$ & entry of $\tQ$\\
        $\etQ_{i,s,a}$ & $q$-value for agent $i$ recommendation $s$ action $a$ \\
        $\mQ$ & $q$-table of a single agent, a slice of $\tQ$, $\tQ_{i,:,:}$ \\
        $\sQ\textbf{SPACE}$ & space of all $q$-values
    \end{tabular}
    \caption{Nomenclature and symbol glossary.}
    \label{tab:nomenclature}
\end{table}

\subsection{Rewards, States, and Actions}\label{sec:rl}

The policies of $Q$-learners playing a repeated congestion game induce a distribution over actions $P_t(\rva)$ with domain $\mathcal{A}^{n}$ at each iteration $t$ where $\rva$ is the action profile specifying an action for each agent. When the $Q$-learners receive recommendations $\vs_t$, the distribution depends on $\vs_t$ which we denote as $P_{\vs_t,t}(\rva)$. As all agents are treated to be equal (un-weighted congestion game) the congestion in the network is determined by mapping the distribution $P_{\vs_t,t}(\rva)$ over action profiles to a simpler distribution over actions $P_{\vs_t,t}(\ra)$ with domain $\mathcal{A}$ instead of $\mathcal{A}^n$. Then, we formulate the goal of the RS as trying to get as close as possible to the target distribution $P^*(\ra)$ which maximizes the social welfare of the congestion game. The optimal target $P^*(\ra)$ is the solution to a traffic assignment problem \cite{boyles2020transportation}. Define the cumulative discounted sum of future rewards as $G = \sum_{t=1}^\tau \gamma^t r_t$ for a given discounting factor $\gamma$. Then, the rewards $r_t$ are equal to the Kullback--Leibler divergence between the induced distribution and the target distribution:

\begin{equation}
\label{eq:reward}
    r_t = - \KL \left( P_{\vs_t, t}(\ra) \Vert P^*(\ra)\right).
\end{equation}

We denote the reward function for the RS as $R_{RS}$. The goal of the RS is to find the policy $\pi^*_{RS}$ which outputs recommendations for the $Q$-learners:
\begin{align*}
    \pi^{*}_{RS} & = \max_{\pi_{RS}} \E_{\sim \pi, \pi_{RS}} [\: G \:] \\
    & = \max_{\pi_{RS}} \E\bigg[- \sum_{t=1}^\tau \gamma^t \KL \left( P_{\vs_t,t}(\ra) \Vert P^*(\ra)\right)\bigg]
\end{align*}

We assume the RS has a state space that includes the $q$-values of the users. For each agent, $Q:\mathcal{S} \times \mathcal{A} \rightarrow \R$, and given $n$ agents, the space of all $q$-values is $\sQ\textbf{SPACE} = \R^{|\mathcal{S} \times \mathcal{A}| n}$. Thus, the RS policy maps $\sQ\textbf{SPACE}$ to $\mathcal{S}$, $\pi_{RS}: \sQ\textbf{SPACE} \rightarrow \mathcal{S}$.

\begin{figure*}[!h]
    \centering
    \includegraphics[width=0.8\textwidth]{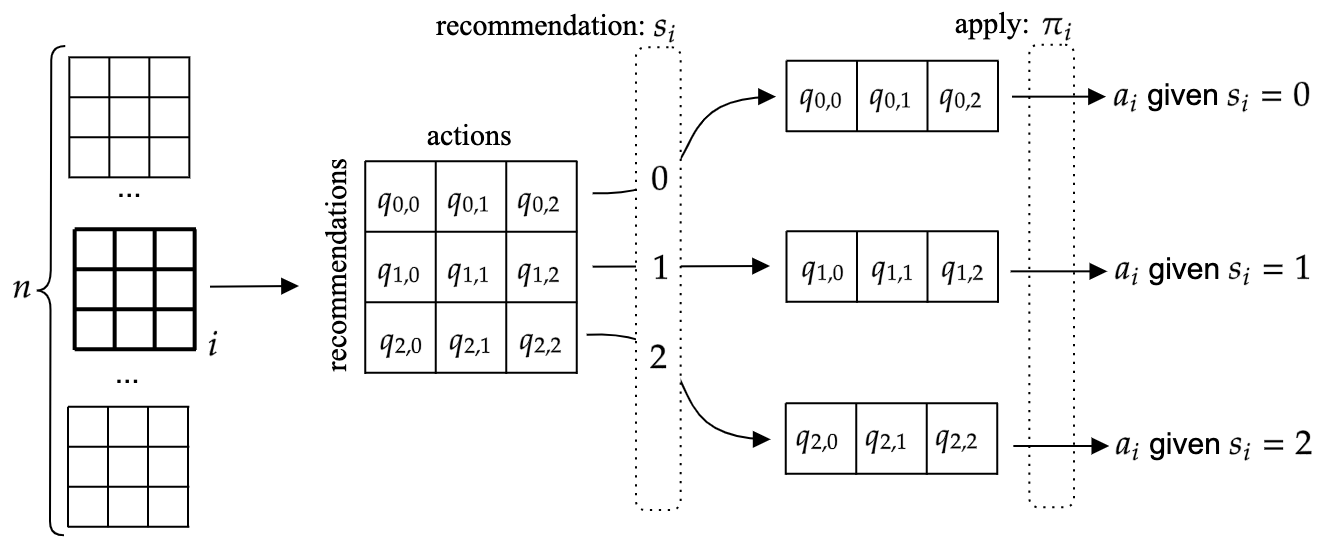}
    \caption{As each user learns a $q$-table of recommendation--action pairs, the recommender chooses the row of the $q$-table by recommending $s_i$ which determines the row to which the policy $\pi_i$ is applied. Time subscripts are omitted for clarity. This is a particular case where the number of actions equals the number of recommendations ($k=m$).}
    \label{fig:recommender_diagram}
\end{figure*}

\subsection{States as Recommendations}\label{sec:states}
The essential modelling assumption of our recommendations is that the $Q$-learners learn the values of actions $a\in\mathcal{A}$ as a function of the recommendation states. The RS outputs a length $n$ vector $\mathbf{s_t} = (s_{1,t}, \dots, s_{n,t})$ from a finite recommendation space $\vs_{i,t} \in \mathcal{S} = \{1, 2, ..., m\}$ and at each timestep provides each $Q$-learner $i$ with a recommendation $s_{i,t}$. The $Q$-learners receive this recommendation and select actions using their $\epsilon$-greedy policies $\pi_i(s_{i,t})=a_{i,t}$. This process is visualized in \autoref{fig:recommender_diagram} for the case where $Q$-learners have 3 actions ($k=|\mathcal{A}|=3$) and the recommendation space has 3 elements ($m=|\mathcal{S}|=3$). The RS is effectively picking the rows of the $q$-tables that $Q$-learners consider when applying their policy.

Our modelling assumption leads to a system where $Q$-learners learn what to do best in response to a recommendation. The $Q$-learners do not prioritize the recommendation over other actions. The recommendation is simply a state which the $Q$-learners use in their learning during their updating. $Q$-learners may learn to pick routes which do not correspond to the recommendation. Likewise, a RS may recommend a route which the $Q$-learner has learned not to follow. Therefore, the ability to influence the outcome of the congestion game is coupled with the learning dynamics of the $Q$-learners.

\subsection{Scaling}\label{sec:scaling}

The RS picks a recommendation $s_i \in \mathcal{S}$ for each agent $i$. Given a recommendation space of size $m$ the action space of the RS becomes $\mathcal{S}^{n}$ and thus has size $|\mathcal{S}^{n}|= m^n$. In any practical application of $Q$-learning running on computers with finite precision, $\sQ\textbf{SPACE}$ will be finite. Given a finite action space, the system becomes an MDP defined by the tuple $(\sQ\textbf{SPACE}, \mathcal{A}_{RS}, T, R_{RS}, \gamma)$, where $T$ is the transition probability distribution between states $\vq\in \sQ\textbf{SPACE}$ given recommendations $s$, $T(\vq'| \vs, \vq)$. For MDPs there are polynomial-time algorithms to find the optimal policy \cite{papadimitriou1987complexity}. However, this MDP grows very fast. $\sQ\textbf{SPACE}$ grows exponential in $n,m,k$: $\mathcal{O}(C^{nmk})$. The action space $\mathcal{A}_{RS}$ grows exponential in $n$ and polynomial in $m$: $\mathcal{O}(m^n)$. However, the following arguments do not require the MDP assumption and give a sense of the extent to which a RS can influence the learning dynamics of systems of $Q$ learners in congestion games.

\subsection{Demonstrating how recommendations can steer learning dynamics}\label{sec:proofs}

We wish to demonstrate that state recommendations can \textit{in theory} steer the learning dynamics. In so doing the RS acts as a controller of a non-linear discrete-time dynamical system. For non-linear discrete-time dynamical systems we can not guarantee \textit{controllability} in a control theoretic sense \cite{liu2016control}. Therefore we will resort to demonstrate some weaker notions than controllability, namely that under given assumptions it is possible to influence the actions that agents take. We then provide intuitions that can aid in understanding the design principles of our heuristic algorithm. In \autoref{sec:theorems} we prove that increasing the size of the recommendation space can increase the steering potential. In \autoref{sec:belief} we show that it is possible to pick recommendations to optimize belief updates. In \autoref{sec:constant} we show that it is necessary to vary recommendations in time to steer the system.

\subsubsection{Steering by increasing the recommendation space}\label{sec:theorems}

We will first present all of our arguments for the single agent case, which are then naturally extended to all agents to ensure that the entire dynamical system can be influenced. Proofs are in \autoref{appendixProof}. Consider a single agent with an $m\times k$ $q$-table $\mQ$ and a deterministic policy $\pi(s) = \argmax_a \mQ_{s,a}$. \begin{definition}
    (\textbf{Reachable}) We define the reachable set of an agent as the set of all actions that maximize the row $j$ of $\mQ$.
    \begin{equation}
        \mathcal{R}(\mQ) = \{a \in \mathcal{A} | \exists_{s} \: \pi(s) = a \}. 
    \end{equation}
\end{definition} Our given $\mQ$ is a $q$-table with $m$ rows which can be extended to have more rows. We define a new function $\mathit{Ext}$ which extends a matrix $\mQ$ to a matrix $\mQ'$ which has $m+1$ rows, and the first $m$ rows are identical to the rows of matrix $\mQ$. $\mathit{Ext}$ is a function $\mathit{Ext}: (m\times k)\text{-matrices} \rightarrow (m+1\times k) \text{-matrices}$ so that $\mQ' = \mathit{Ext}(\mQ)$ and $\forall$ indices $i,j$ where $1\leq i \leq m, 1\leq j \leq k$: $\mQ'_{i,j} = \mQ_{i,j}$. With the newly defined tools we can state our first theorem.
\begin{theorem}\label{theorem:1}
    (\textbf{Increasing Reachability}) When increasing the size of the recommendation space $m$ which amounts to adding rows to a $q$-table $\mQ$, the size of the reachable set $R(\mQ)$ is monotonically non-decreasing.
    \begin{equation*}
        \mathcal{R}(\mQ) \subseteq \mathcal{R}(\mathit{Ext}(\mQ))
    \end{equation*}
\end{theorem} The function $\mathit{Ext}$ can be seen as adding a new row which is a vector $\vv=(q_1, ..., q_k)$ of $q$-values.
\begin{assumption}\label{assumption} (\textbf{Full Support})
The $q$-values of the new row $\vv$ are drawn from a distribution $D$ where each action $a$ has a non-zero probability of being the $\argmax$ of $\vv$: 
$$\vv \sim D \text{ and } \forall a, P(\argmax(v) = a) > 0.$$
\end{assumption}
\begin{theorem}\label{theorem:2} (\textbf{Global Reachability}) As the size of the recommendation space approaches infinity, and given Full Support of the $q$-values, the RS can induce any action in the $Q$-learner:
    \begin{equation*}
        \lim_{m\rightarrow \infty} \mathcal{R}(\mathit{Ext}^{m}(\mQ)) = \mathcal{A}.
    \end{equation*}
\end{theorem} With these two theorems, \textbf{Increasing Reachability} and \textbf{Global Reachability}, we have shown that the size of the recommendation space has a big influence on the states that the RS can induce for a single $Q$-learner. If we have a system of $n$ $Q$-learners playing a game, all of our results hold for the joint system too, with a tensor composed of all the $q$-tables of agents, $\tQ = (\mQ_1, ..., \mQ_n)$, and the reachable set as the set $\mathcal{R}(\tQ) \subseteq \mathcal{A}^n$ of action profiles $\va$. It will still hold that increasing the size of the recommendation space for any of the $Q$-learners, and thus the number of rows in their $q$-tables, we can increase the size of the reachable set for the system.
\begin{definition}
    (\textbf{Steering Potential}) Given a RS which can strategically pick recommendations $\vs$, and users described by $q$-tables $\tQ$ and $\argmax$ policies, we define the steering potential of the RS as $\mathcal{R}(\tQ)/\mathcal{A}$.
\end{definition} The steering potential reflects the proportion of possible action profiles which a recommender can induce. It follows from the definition that maximizing reachability maximizes the steering potential. Fast forward to \autoref{sec:results2}, increasing the size of the recommendation set allows our algorithm to drive the system to near-system-optimum with finitely many states. In the next section we explain the other design principles required to achieve this performance.

\subsubsection{Steering By Optimizing Belief Updates}\label{sec:belief}

Optimizing social welfare requires \textit{optimizing belief updates}. This is done to achieve a population of $Q$-learners that believes the socially optimal actions to be better than the socially sub-optimal actions. Doing so for the Braess Paradox amounts, practically, to `tricking' $Q$-learners into updating their $q$-values such that they believe "cross" to be a worse action than it actually is. This is challenging because "cross" always appears as the optimal greedy action.

\textbf{Example} Consider a single user, and a single $q$-table  \\
$\mQ = 
\begin{blockarray}{cccc}
a_1 & a_2 & a_3 \\
\begin{block}{(ccc)c}
  q_{1,1} & *q_{1,2} & q_{1,3} & s_1 \\
  q_{2,1} & *q_{2,2} & q_{2,3} & s_2 \\
\end{block}
\end{blockarray}$
for actions $(a_1, a_2, a_3)$ \\
and recommendations $(s_1, s_2)$.

We use an asterix $*$ to indicate the $\max$ values of the row. Both $q_{1,2}, q_{2,2}$ $\max$ values belong to the same column, $a_2$, so for both recommendations a $Q$-learner with an $\argmax$ policy would pick action $2$. We know that the $Q$-learner's $q$-values change according to the Bellman update rule from time $t$ to time $t+1$, and can express the total update as:

\begin{equation}
    \Delta_{s,a} = q^{t+1}_{s, a} - q^{t}_{s, a} =  \alpha\bigg[r + \gamma\max_{a'}q^{t}_{s', a'} - q^{t}_{s, a}\bigg].
\end{equation}

The reward $r$ specified the specific reward for agent $i$ (though subscripts are omitted). Given $\vs$, $r_i$ is determined:

\begin{equation}
    r_i = R(\pi_1(s_1), \pi_1(s_2), \dots, \pi_1(s_n)).
\end{equation}

Then recommendations can be picked to optimize belief updates. Let us suppose that $a_2$ is a socially beneficial action. Given that it is the $\argmax$ of both of the recommendation states, we can interpret this agent as having learned socially beneficial behavior for both recommendations. If we wish to ensure that these socially beneficial actions will stay in the reachable set $\mathcal{R}(Q)$, we should use \textbf{positive reinforcement} and select $\argmax_{s}\Delta_{s,2}$. If instead action 2 is socially sub-optimal, we should use \textbf{negative reinforcement} and select $\argmin_{s}\Delta_{s,2}$.

\subsubsection{Constant Recommendations Change Nothing}\label{sec:constant}

We define a recommendation as constant when it does not change from the perspective of the user. Thus, for a constant recommendation, each user may receive a different recommendation, so long as the recommendation remains unchanged across iterations. If the RS picks a constant recommendation vector $\mathbf{s}$ and never alters it during repeated play of the congestion game, the learning dynamics will resemble the case without recommendations, and collapse to the stateless congestion game as described in \autoref{sec:stateless}. This is the case for any arbitrary recommendation vector, and means that static no-regret learning formalisms \cite{gordon2008no} are not directly meaningful and applicable. To see this clearly, consider \autoref{fig:recommender_diagram} and imagine what happens when the same recommendation is picked for the entire horizon: the same $q$-values are used to select actions and update, which leads the system dynamics to be identical to the case where there are no recommendations.

\textbf{Example} Take a new $\mQ = \begin{blockarray}{cccc}
a_1 & a_2 & a_3 \\
\begin{block}{(ccc)c}
  *q_{1,1} & q_{1,2} & q_{1,3} & s_1 \\
  q_{2,1} & *q_{2,2} & q_{2,3} & s_2 \\
\end{block}
\end{blockarray}$, and assume $a_1$ is the socially beneficial action. It is the case that $a_1\in\mathcal{R}(\mQ)$, so we can induce action 1 by recommending $s_1$. However, after the $Q$-learner updates the $q$-table to $\mQ'$, there is no guarantee that $a_1\in\mathcal{R}(\mQ')$. Specifically: 
$$a_1 \not\in\mathcal{R}(\mQ') \text{ if } q_{1,1}+\Delta_{1,1}<\max\{q_{1,2}, q_{1,3}\}$$ and the update $\Delta_{1,1}$ will have altered the $\argmax$.

Therefore, a crucially relevant feature of the LDMP is that the RS must vary the recommendations throughout the learning process to steer the learning dynamics. A constant recommendation vector induces no perceivable change to the learning dynamics. This feature can be understood from physics as a dynamical system which is being forced. If the forcing function is constant, the physical system may be translated, but its behavior remains symmetric with the un-forced system. On the other hand, if the forcing function is time-varying like a sinusoid with a particular frequency, then the dynamical system can be forced into a resonant mode which creates behavior which is no longer symmetric with the un-forced system.

\section{Results}\label{sec:results2}

In this section we use our proposed recommendation algorithm to control the learning dynamics of $Q$-learners in the Braess Paradox. Our notion of recommendations generalizes to include cases where the number of available recommendations far exceeds the possible routes. In other words, the RS can be thought of as providing signals to the $Q$-learners which are no longer directly associated with specific routes. Instead, these signals become a kind of correlating tool which the RS has access to and can use to steer the users. In fact, we demonstrate that our algorithm algorithm becomes more effective as the number of available recommendations ($m$) increases. While this section focuses on values of $m\geq3$, we study the case where recommendations and actions align ($m=3$) in great detail in \autoref{appendixRouteRecommender}.

To showcase the effectiveness of the proposed recommender algorithm\footnote{Code available at: \url{https://anonymous.4open.science/r/manipulateRL-57DB}. For a detailed description of the algorithm please refer to \autoref{appendixPseudocode}.} we run $\epsilon$-greedy $Q$-learners ($\alpha=0.1, \gamma=0$) in the Braess Paradox and analyze the social welfare of the system over a fixed horizon of $10000$ learning steps. In these experiments we will decay the exploration rate $\epsilon$ of the $Q$-learners to $0$ over time. On the one hand, this forces the $Q$-learners to converge so that we may study their converged behavior. On the other, this gives us a chance to study a scenario where the model the RS has of the $Q$-learners policies improves over time, since the error rate of the RS's model is equal to $\epsilon$. We vary the number of $Q$-learners from 100 and 900 to demonstrate that our algorithm can scale to many users. We also vary the size of the recommendation space from 3 to 93 to demonstrate that our algorithm is more effective with a larger recommendation space, as proven in \autoref{sec:theorems}. To benchmark our algorithm (optimized) we compare it with (none) no recommendations and random recommendations. Each setting--population--state combination is run 40 times and the reported values are averaged over these runs. 

\paragraph{Mapping latency to Social Welfare}
All subsequent results are displayed in terms of the social welfare, which is a normalized average latency of the network. In the augmented network of the Braess' Paradox a social welfare of 0 corresponds to the worst-case congestion achieved at the Nash Equilibrium where all agents cross for an average latency of $2$. At this level, the Price of Anarchy is highest. On the other hand, a social welfare of 1 corresponds to the best-case congestion at the socially optimal utilization where half of the agents pick $u$ and half of the agents pick $d$ with an average latency of $1.5$. At this level, the Price of Anarchy is minimized.

\begin{figure}[h!]
    \centering
    \includegraphics[width=\linewidth]{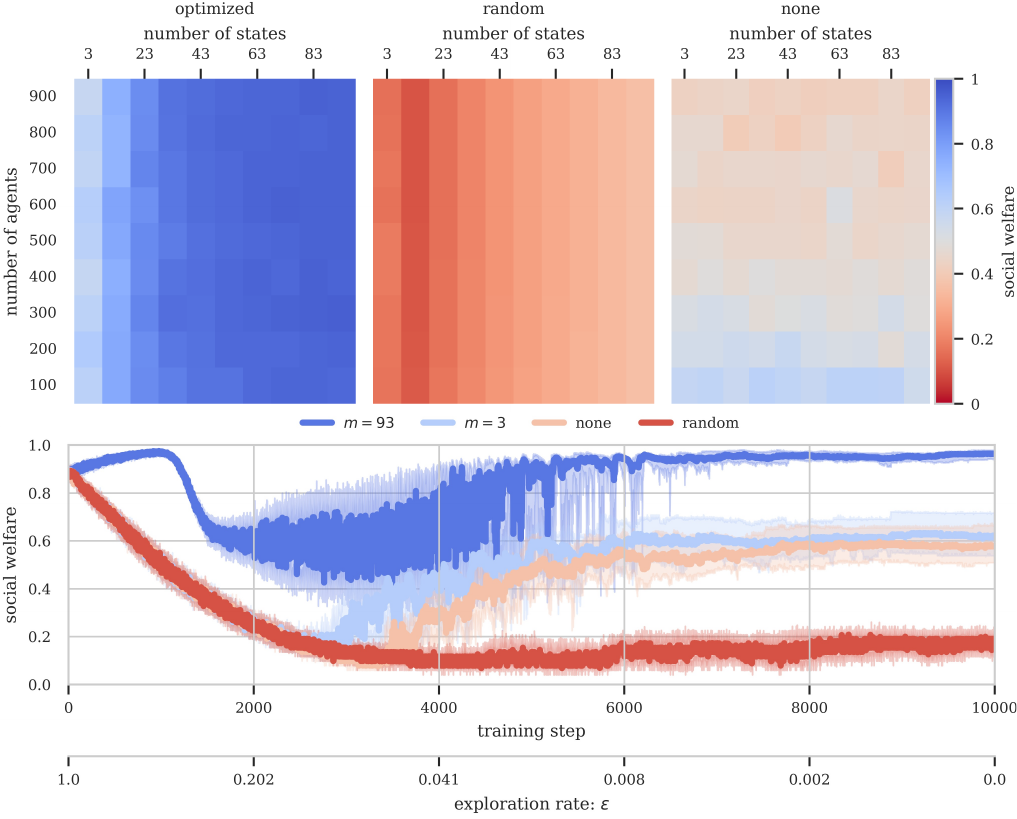}
    \caption{\textit{Top:} Social welfare achieved in Braess's Paradox while varying, the numbers of $Q$-learners, the size of the recommendation space, and the type of recommender: optimized, random, and none. \textit{Botttom:} Evolution of the social welfare for four select conditions. Values of social welfare were rescaled from $[-2,-1.5]\rightarrow[0,1]$, higher social welfare is better.}
    \label{fig:results}
\end{figure}

In \autoref{fig:results} we report the results of our experiments. The heatmaps presented at the top represent the average social welfare achieved over a finite horizon by each run. We note that the heuristic recommender setting achieves consistently higher values of social welfare (than the \textit{random} or \textit{none} settings) regardless of the number of agents. The setting without recommendations achieves better welfare than \textit{random} recommendations which consistently give the worst results. Furthermore, we note that increasing the number of states for the recommender setting increases the social welfare of the system. At the bottom of \autoref{fig:results} we report the learning dynamics in terms of the social welfare (over the training steps and decreasing exploration rate $\epsilon$) for four conditions. The system with random recommendations with the least number of recommendation states ($m=3$) converges to the NE. On the other hand, the system with no recommendations (\textit{none}) converges to a better value than NE, but fails to reach the social optimum. 
In contrast, using the heuristic algorithm to recommend the optimized states shows some improvements.
On one extreme end, for the system with the least recommendation states ($m=3$), we reach social welfare that is comparable to \textit{none}, but with slightly faster convergence.
% The system with recommendations with small number of recommendation states converges slightly faster than the system with no recommendation but reaches comparable social welfare. 
But, on the other extreme of more recommendation states ($m=93$), recommendations can steer the system performance to converge very close to the social optimum. It is worth noting that all systems exhibit a transient in the early phases of training (when the $\epsilon$ value is high).

\begin{figure}
    \centering
    \includegraphics[width=\linewidth]{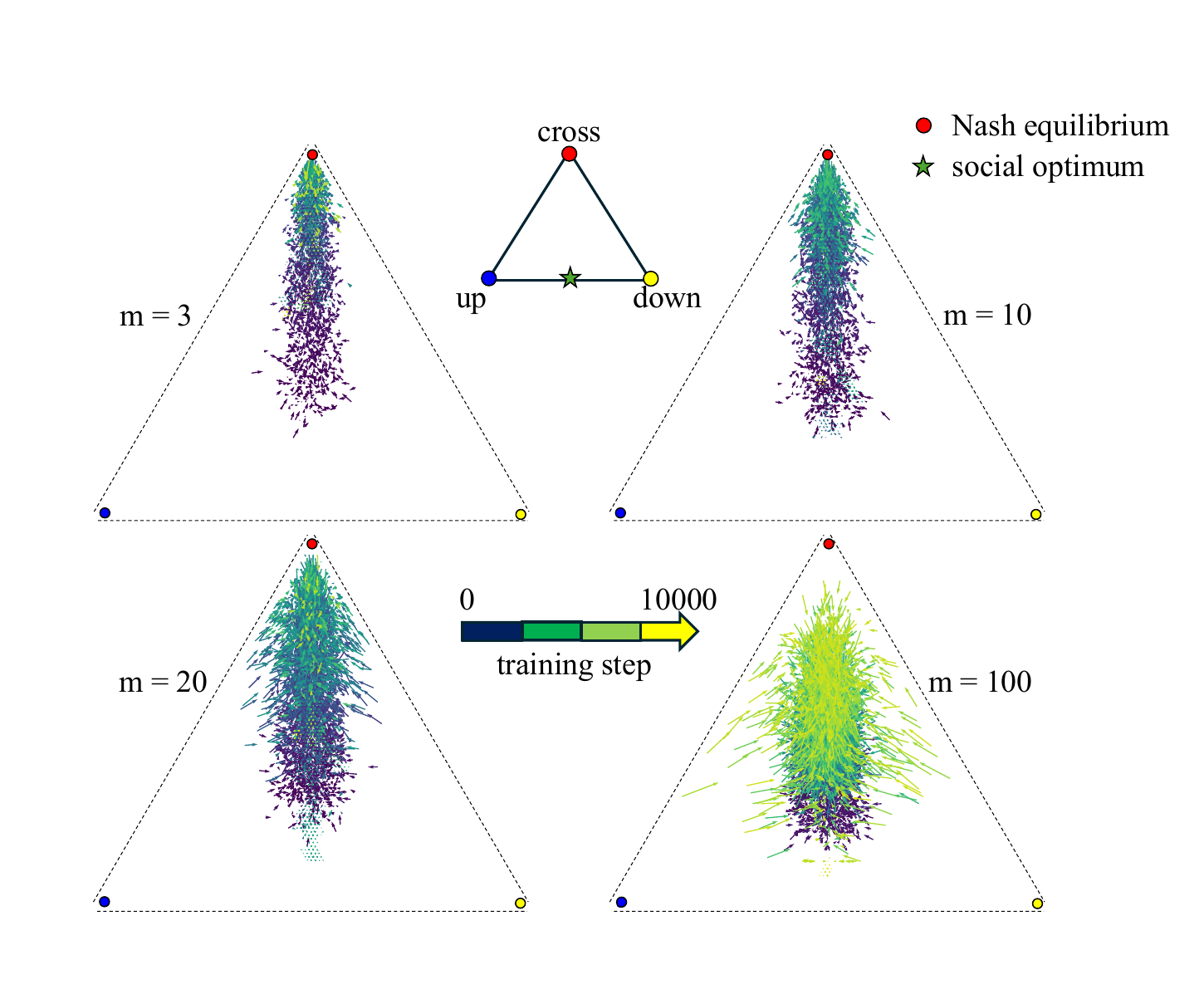}
    \caption{Vector plots to visualize the distributions and change of actions during the training steps while $Q$-learners receive optimized recommendations. The corners of the triangles indicate the points where all agents pick the same action (top: cross, left: up, right: down). Vectors point from the action profile of training step $t$ towards the action profile of training step $t+1$. Training steps are color-coded to increase from dark blue ($t=0$) to yellow ($t=10000$). We can observe steering as $m$ increases, from $3$ to $100$, as vectors are shifted away from the Nash Equilibrium (top) towards the social optimum (midpoint of the left and right corners).}
    \label{fig:steering}
\end{figure}

\autoref{fig:steering} shows in detail how the recommender algorithm affects the selection of actions during learning. The triangle plots compactly represent all possible action profiles of the agents, where each corner is the action profile where all agents pick the same action. It is clear that as the number of possible recommendations $m$ increases, the action profiles are steered away from the Nash Equilibrium and towards the social optimum.

\section{Discussion and Conclusion}\label{sec:discussion}

In this paper, we have formalized the Learning Dynamic Manipulation Problem (LDMP) which we propose as a model of route recommendations which take two main dynamics into account: the effects on the learning of users, and consequently the effects on the social welfare. We show that our model has a \textit{steering potential} which increases with the size of the recommendation space. Furthermore, we designed a heuristic algorithm and demonstrated in the Braess Paradox that we can steer the learning dynamics of a system of $Q$-learners ($Q$-learners). We interpret the success that any RS may have in such a system as due to its ability to i) slow down the dynamic evolution of the system which naturally ``degrades'' to the NE, and ii) speed up the dynamic evolution of the system, which, close to the NE, experiences a repelling force (\autoref{fig:braess}, right). \autoref{fig:results} also confirms our intuitions from \autoref{theorem:1} and \autoref{theorem:2} that the steering potential increases with the size of the recommendation space (number of states). This result could be interpreted as more information allowing for better social outcomes.

In \autoref{sec:results2} we generalize our recommendations to the notion of ``signal" that does not correspond to a particular action, but is rather a piece of correlating information. One could imagine that drivers in traffic receive colors as signals. Our model then assumes that these colors are treated as meaningful learning information and used to update $q$-values. In \autoref{sec:results2} we experiment with this notion and use our heuristic algorithm to show that the welfare-improving effects are enhanced as the ratio of signals to actions grows: in other words, as the RS can use more signals which the $Q$-learners will find meaningful to their learning.

This model, however, allows for other interpretations. While in the presented example, the RS is directed at optimizing social welfare, nothing prevents the RS from steering towards other metrics. Simultaneously, the model could also be seen as `confusing' the $Q$-learners which leads to a more grim interpretation that confusing the learning dynamics of users allows for greater manipulation. Such an interpretation may have interesting and relevant extrapolations to the present world where digital systems bombard their users with large amounts of information. For these reasons, we emphasize that this formalism should be as alarming as it may be insightful. In \autoref{sec:rec_aling} we include additional results and discussions about recommendation alignment and what it means for a route RS to be aligned with its users.

% The heuristic recommender we propose can be generally applied to any congestion game. On the one hand, the LDMP formalism guarantees the existence of an optimal policy for an RS that knows the dynamic beliefs of users. While it is true that the recommender is unlikely to have a perfect knowledge of the beliefs of the users, it would likely be able to model them to a sufficient extent. Importantly, this formalism provides tools to study and theorize about the interaction and effects of RS and their users. Furthermore, the proposed algorithm offers a baseline for potential future developments, such for example a recommender that is (like the users) a $Q$-learner. We discuss next some of these considerations.

% % \paragraph{Interpreting the model}
% What does it mean for recommendations to be received as states? For $Q$-learners it means that each recommendation is treated as meaningful information on top of which to learn.

% On the other, the system reaches a better performance due to what could also be seen as confusion of the agents, who are kept from realizing that there is an action that is more beneficial to them (at least in the short term). Thus the increase of information entropy would seem to lead to greater confusion, which could be likened to today's media of the infosphere bombarding its users with so much information that they become confused and easier to manipulate. 

Our results have a notable limitation that we assume that the RS is omniscient, somewhat ``God-like'', such that the $q$-tables of all agents were known and the welfare optimizing actions for all agents were also known. This omniscience is a very restrictive assumption that will not apply in practice. Nonetheless, we wish to point out that while omniscience will likely not hold, RSs in practice all create models of user behavior. While predictive power on an individual scale is harder, aggregating RS users into categories has been a successful means for RSs to model users, as in collaborative filtering \cite{su2009survey}. Therefore, even though it is unrealistic to assume omniscience, it is still interesting to consider what omniscience can afford while keeping in mind that omniscience is the best-case scenario of any RS. In \autoref{sec:results1} we run some simulations with relaxed assumptions of omniscience by adding noise to the $Q$-learning dynamics.

% We also considered cases that partially relax the assumption of omniscience by assuming that the users use an $\epsilon$-greedy policy rather than only an $\textbf{argmax}$ on their $q$-values.
% This additional source of noise which the RS was not aware of meant that predicting the actions of users from their $q$-values was not deterministic.
% Increasing the exploration rate $\epsilon$ of the $\epsilon$-greedy $Q$-learning users introduced more noise and reduced the predictability of the effects of recommendations, rendering manipulation ineffective and hindering learning the social optimum (\autoref{fig:recommender_results}).

% Furthermore, we showed that when the beliefs of agents are known, the problem reduces to an MDP, but when the beliefs are not known, the problem becomes a POMDP. While an MDP admits an efficiently computable optimal policy, a POMDP does not, and finding the optimal policy is worst-case complexity \textbf{NP}-hard. Given the unlikely nature of the omniscience of an RS, we can conclude that the Learning Dynamic Manipulation Problem does not have an efficiently computable optimal policy for practical cases. Nonetheless, there may exist practical approximations.

% \paragraph{Ethical considerations} 

Another limitation of the work is that we run our experiments on the Braess Paradox, which is admittedly small. However, the Braess Paradox is a widely studied feature of complex networks which has found many applications \cite{fisk1981empirical, arnott1994economics, steinberg1983prevalence, christodoulou2005price, di2014braess, zhao2014braess, acemoglu2018informational}. Furthermore, it is not immediately clear how to extend this network to larger sizes. As our work deals with recommendations we believe the crucial dimension of scale is the number of agents, which represent people to which one can recommend. It is the population scaling that is especially important in practice (if we can only recommend to a few users effectively we would lose customers). Though it is known that Braess-like features can exist in larger networks \cite{valiant2006braess, chung2010braess}, the Braess paradox we use is an abstract representation of a particularly challenging coordination problem and extending the network on which we define it does not add much to the analysis. In practice, in traffic, one could be repeatedly confronted with the Braess paradox when going through a given urban area.

% \paragraph{Future work} 
In future work we are interested to better understand the \textit{controllability} of multi-agent reinforcement learning systems both from an algorithmic lens, and as models of learning behavior. In this pursuit, we are interested to understand the proposed formalism in the context of other games. This sort of dynamic could also occur in RSs that provide content recommendations \cite{ko2022survey}. Such systems also create feedback loops with their users (never recommending certain content to a group of users, only showing users what they want to see, etc.), which can drive content bubbles and echo chambers. We propose that our approach may be used for RSs to optimize a notion of social welfare, while also achieving the necessary goals of a platform e.g. engagement. Both route RSs and content RSs need to account for the effects of their actions on the population they are affecting in order to consistently maintain high level of recommendations. Finally, future work should test the limits of dynamic programming with a large Neural Network to solve the LDMP. We encourage using our heuristic approach as a benchmark for black-box methods.

\section{Impact Statement} 
This paper presents work whose goal is to advance the understanding of the effects that route recommendations have on social welfare. To do so we developed an algorithm which could steer the learning dynamics of algorithmic users. While we feel this to be a model which is not directly applicable to humans, we do recognize the potential negative impacts, as many others have, of recommendations on human users. Therefore, we hope this paper to be taken as a necessary step in our understanding of these systems, such that we can curb the potential dangers of algorithms recommending things to human users.


\section*{Acknowledgments}
The authors thank Dirk Helbing and Heinrich Nax for their useful comments and discussions concerning the manuscript, as well as the anonymous reviewers for their clarifying reviews and suggestions. This project received funding from the European Union (ERC, CoCi, 833168).

\bibliographystyle{elsarticle-num}
\bibliography{references}

\newpage
\appendix
\onecolumn

\section{Learning Dynamics on the Braess Augmented Network}\label{appendixA}

\renewcommand{\thefigure}{S\arabic{figure}}
\setcounter{figure}{0}

% \begin{figure*}[!ht]
% \centering
% \begin{subfigure}{.5\textwidth}
%   \centering
%   \includegraphics[width=0.8\linewidth]{figures/recommender/braessNetworkInitial.pdf}
%   \caption{Initial Network}
%   \label{fig:sub1}
% \end{subfigure}%
% \begin{subfigure}{.5\textwidth}
%   \centering
%   \includegraphics[width=0.8\linewidth]{figures/recommender/braessNetwork.pdf}
%   \caption{Augmented Network}
%   \label{fig:sub2}
% \end{subfigure}
% \caption{Illustration of the initial network (a), and the augmented network (b) in the Braess Paradox. Agents start in the ``S'' state and pick a path to reach state ``t''. The numbers represent the cost of traveling over a link. A cost of $x$ is the ratio of agents that choose that link. Two actions are possible in (a), $\mathit{up}$ takes the upper edges, and $\mathit{down}$ takes the lower edges. In (b) an additional action $\mathit{cross}$ is possible, which takes the first upper edge, crosses to the lower section at the middle, and finishes on the second lower edge. Rational and fully-informed agents all pick the crossing link in the augmented network (Nash equilibrium), which leads to high congestion and the worst possible social welfare.}
% \label{fig:braess_network_appendix}
% \end{figure*}

\begin{figure*}[!ht]
    \centering
    \includegraphics[width=0.8\textwidth]{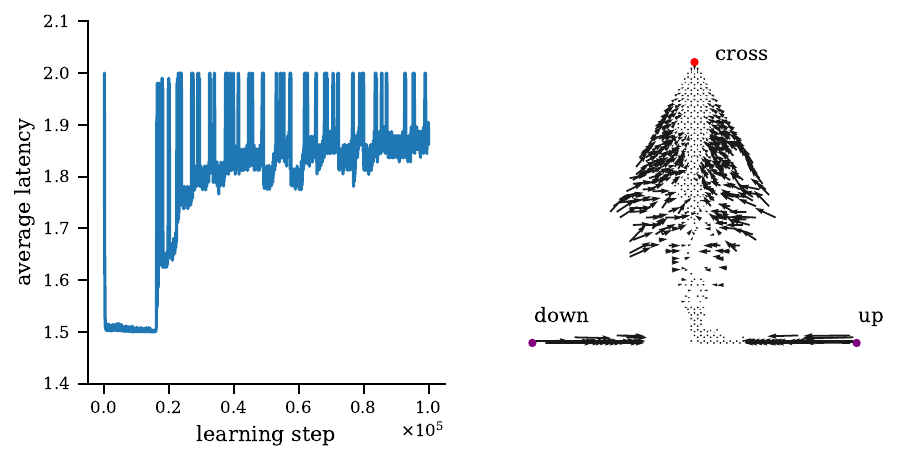}
    \caption{\textit{Left:} The latency that emerges from the actions of the learning agents exhibits chaotic dynamics that tend towards the Nash equilibrium with $\mathbb{E}[l]=2$. \textit{Right}: Vector field visualization of the evolution of the joint action space of the agents $(n_{up}, n_{down}, n_{cross})$. The corners of the triangle represent all agents taking a single action, and the arrows indicate the change between consecutive learning steps. This work is conducted in detail in \cite{carissimo2024counter}.}
    \label{fig:braess}
\end{figure*}

\section{Theorem Proofs}
\label{appendixProof}

\subsection{Proof of \autoref{theorem:1}}
\begin{proof}
    Consider matrices $\mQ$ and $\mQ' = \mathit{Ext}(\mQ)$. We can split the $\mQ'$ matrix into $\mQ'_{1:m,1:j}$ and $\mQ'_{m:m+1,1:j}$ (the row added in the $\mathit{Ext}$ operation). Since the $\mathcal{R}$ operates on rows independently and is associative we have $\mathcal{R}(\mQ') = \mathcal{R}(\mQ'_{1:m,1:j}) \cup \mathcal{R}(\mQ'_{m:m+1,1:j})$. Since by definition of $\mathit{Ext}$ we have $\mQ = \mQ'_{1:m,1:j}$ we arrive at $\mathcal{R}(\mQ') = \mathcal{R}(\mQ) \cup \mathcal{R}(\mQ'_{m:m+1,1:j})$. Therefore $\mathcal{R}(\mQ) \subseteq \mathcal{R}(\mathit{Ext}(\mQ))$.
\end{proof}

\subsection{Proof of \autoref{theorem:2}}
\begin{proof}
    Take an arbitrary action $a\in {1, \dots, k}$ which has probability $p > 0$ of being the $\argmax$ of every new added row. At any recommendation space size $m$ the probability that $a$ never appears as the $\argmax$ of any row can be expressed as: $(1-p)^{m}$. As $m\rightarrow\infty$, the probability $(1-p)^m$ goes to $0$. Therefore, with probability $1$, the action $a$ will be the $\argmax$ of at least $1$ row, and the $a$ can be induced in an agent.
\end{proof}

\section{Heuristic Algorithm Pseudocode and Subroutines}\label{appendixPseudocode}

\autoref{alg:cap} describes a routine that can be used to assign recommendations to $Q$-learners at each timestep of a repeated congestion game. In the algorithm, we call a few subroutines which we explain in greater detail.

The subroutine \Call{EstimateReward}{} calculates an optimistic estimate of the reward for each action $\Bar{\vr}$. It calculates all the agents for which all recommendations only lead to one action. This defines a new congestion game with altered utilities $u'_a$ for each action, and a new socially optimal assignment $d'^{*}$. The reward for each action is estimated as $u_a(d'^{*}_a)$, the new socially optimal actions given the original utility for each action $u_a$.

The subroutine \Call{EstimateUpdate}{} receives the estimated rewards per action and calculates the expected update as a difference $\Delta_{i,s,a}$ between the $q$-values of agents and the estimated reward assuming the $q$-values are updated with the Bellman equation.

The subroutine \Call{CalculatePriority}{} outputs a priority score for each action and reflects the actions that most need agents to be assigned to them. This is done by looking at the difference between the number of unique agents that could be assigned recommendations that lead to $a$, and the optimal value $d^*_a$. Priority 0 means that no more agents need to be assigned to the action.

The subroutine \Call{SelectRecommendation}{} receives a table with actions as columns and rows filled with the values $\Delta_{i,s,a}$ that have been sorted. The routine proceeds row by row, checking first if a column has priority 0, then the column is removed. Second, if the agent $i$ of $\Delta_{i,s,a}$ is unique in the row. If it is, recommendation $s$ will be assigned to agent $i$. If it is not unique then the recommendation is selected for the highest priority action. If the actions have equal priority then the recommendation $s$ is selected such that $\Delta_{i, s, a} < \Delta_{i, s', a'}$.

\begin{algorithm}
\caption{Heuristic algorithm for a route recommender}\label{alg:cap}
\begin{algorithmic}[1]
\Require $\mQ$ \Comment{the $q$-tables of all agents}
\Require $n, k, m$ \Comment{numbers of (agents, actions, recommendations)}
\Require $\vd^*$ \Comment{the target assignment of agents per action}
\For{$i \in \{1, 2, \dots, n\}$}
\State $\mA_i \gets \argmax_a{\mQ_{i,:,a}}$
\State $\mP_{i} \gets \Call{GetPossibleActions}{A_i}$
% $assignActions(a_{p}, R_{e})$
% \If{$\mathrm{len}(\mP_{i}) == 1$}
% \State $\vs_i, \vd_a \gets \Call{AssignAction}{\mP_{i}}$ \Comment{assign unique action $a$ and keep count $\vd_a$}
% \EndIf
\EndFor
\State $\vr \gets \Call{EstimateReward}{\mP}$
% \Comment{indexed by action}
\State $\Delta \gets \Call{EstimateUpdate}{\mQ, \mA, \vr}$ \Comment{a table with update estimates separated by actions}
\State $\mathbf{C_{min}} \gets \Call{SortTable}{C, \min}$
\Comment{sorts to minimize the update estimates by actions}
\State $\mu \gets \Call{CalculatePriority}{\mP, \vd^*}$ \Comment{which actions need agents most}
\State $\vs \gets \Call{SelectRecommendation}{C_{min}, \mu}$ \Comment{assign recommendations to achieve priority}

\If{$\not \forall_{i \in agents} \Call{Assigned}{i}$} \Comment{if agents are unassigned}
\State $\mathbf{C_{max}} \gets \Call{SortTable}{C, \max}$ \Comment{sort to maximize the update estimates}
\State $\mu' \gets \Call{CalculatePriority}{\mP, \vd^*}$
\State $\vs \gets \Call{SelectRecommendation}{C_{max}, \mu'}$
\EndIf

\end{algorithmic}
\end{algorithm}

\section{Three-state recommender experiments: results for other $q$-value initializations}
\label{appendixC}

\begin{figure*}[!ht]
    \centering
    \includegraphics[width=0.8\textwidth]{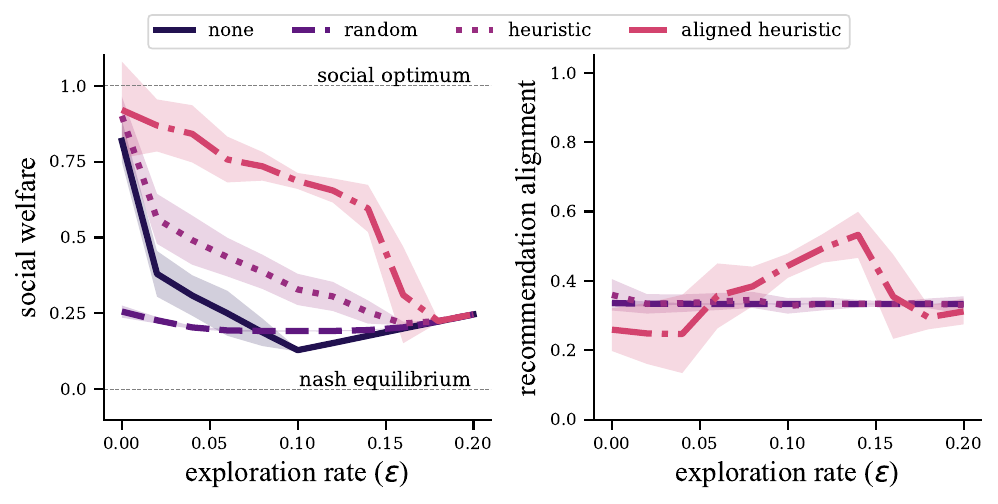}
    \caption{The systems were initialized to uniform random $q$-values.}
\end{figure*}

\begin{figure*}[!ht]
    \centering
    \includegraphics[width=0.8\textwidth]{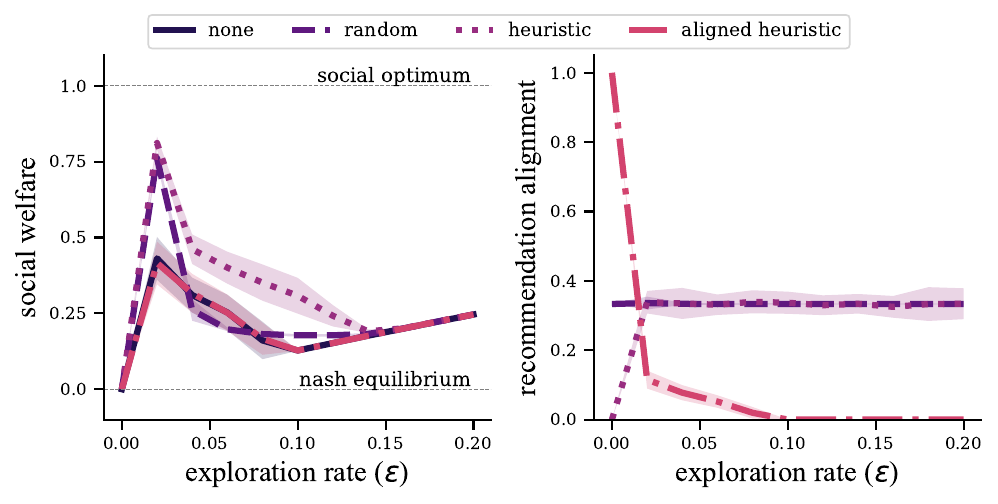}
    \caption{The systems were initialized to the Nash belief $Q=-2$. }
\end{figure*}

\section{Steering in the Initial Network: A simple Learning Dynamic Manipulation Problem}\label{appendixRouteRecommender}

A symmetric routing game is considered with two routes (\autoref{fig:sub1}): $N$ agents ($Q$-learners: $\alpha=0.01$, $\gamma=0$, $\epsilon=0.01$) decide whether to pick up or down, leading to $n_{u}$ agents picking up, and $n_{d}$ agents picking down. The latencies experienced by the agents are linear in the fraction of agents that choose the actions, $\frac{n_{u}}{N}, \frac{n_{d}}{N}$. Specifically, the latencies, $l(a)$, are: $l(u) = 1 + \frac{n_{u}}{N}, l(d) = 1 + \frac{n_{d}}{N}$.

The Nash equilibrium is aligned with the Social Welfare optimizing state, such that the average latency experienced by all agents is $1.5$.
Using the negative of latency as the rewards ($r_i=-l(a_i)$), we choose to initialize all agents equally with the following $q$-values, $Q(\bullet,u)=Q(\bullet,d)= -1.5$.
These are equivalent to the negative of the latency of the actions up and down experienced by agents at the Nash equilibrium (higher latency leads to lower $q$-values). Due to the $\epsilon$-greedy policy, for equal $q$-values an agent picks the first action in their $q$-tables (the definition of the $\mathbf{argmax}$ operator being in line with NumPy). Thus, regardless of the recommendation, the $\mathbf{argmax}$ of all agents will be to go up in the first iteration of the game. The only deviations from this action will be due to the $\epsilon$ exploration rate. This creates a simple scenario for an RS to evaluate the effects of its recommendations on welfare.

\begin{figure}
    \centering \includegraphics[width=\linewidth]{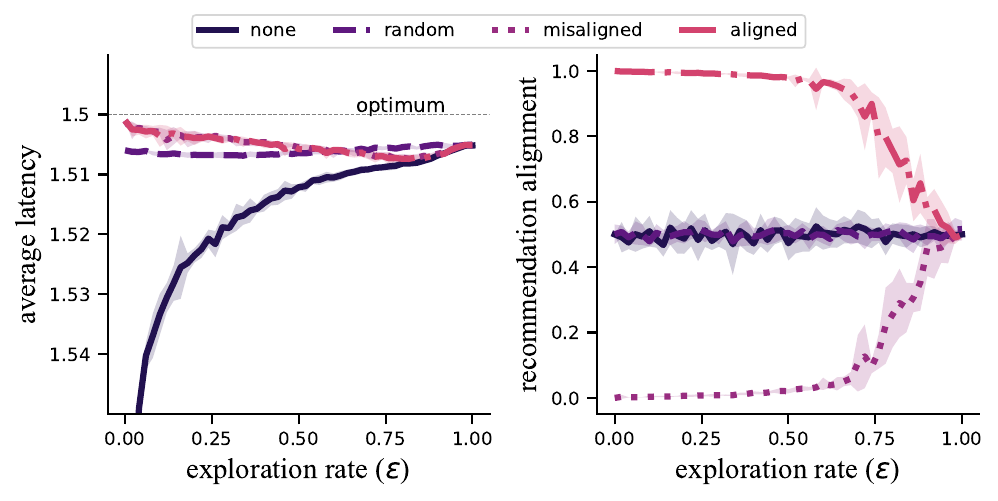}
    \centering
    \caption{100 $Q$-learners ($\alpha=0.01$, $\gamma=0$) are simulated for 500 learning steps in the initial network \autoref{fig:sub1} for varying RS and $\epsilon$ exploration rates. Curves plot averages 10 repetitions of training with error bars as standard deviation. \textit{Left:} Average latency for the different recommender schemes. With the exception of the constant recommender, recommendations can help the users achieve social optimum. \textit{Right:} Alignment (and misalignment) for the heuristic recommenders converge to 0.5 as $\epsilon$ increases to 1 (fully random actions by agents.)
    }
    \label{fig:two_route}
\end{figure}

To demonstrate the potential of the RS, we test three cases in the \autoref{fig:braess_network} scenario, report the results in \autoref{fig:two_route}. (\textbf{none}) This recommendation is kept constant during iterations, and it does not help the agents to converge. In fact, keeping a recommendation constant is equivalent to providing no recommendation. The social welfare is poorest because, due to the $\textbf{argmax}$ definition, agents start all doing the same action, update their beliefs in the same way, and all do the other action in the next iteration. This is only offset by the $\epsilon$ exploration rate. 
(\textbf{random}) The recommendation is randomized between each iteration. The results show significant improvement from constant recommendation. While similar in social welfare to aligned and misaligned recommendations, only half of all recommendations are aligned.
(\textbf{misaligned}) A two-step recommendation, sufficient for agents to immediately converge to the Nash equilibrium. The first recommendation is for all agents to go up. The second recommendation splits the population, as in the fixed case. The recommendation is then kept constant. While achieving rapid convergence to the Nash equilibrium this recommendation is fully misaligned.
(\textbf{aligned}) A two-step recommendation, as in the misaligned case, but the first recommendation is for all agents to go down. Again, the second recommendation splits the population half-and-half and is then kept constant. This recommendation achieves both rapid convergence and recommendation alignment.

These results show that it is possible to steer $Q$-learning convergence with cleverly timed recommendations to induce coordinated behavior in a system of multi-agent $Q$-learners. Additionally, recommendation alignment is measured as a metric that shows whether the coordination was achieved on average with aligned or misaligned recommendations. The difference between the aligned and misaligned cases is minimal but illustrates the effect of picking recommendations on alignment.

\section{Route Recommendation}\label{sec:results1}
\begin{figure}[htb]
    \centering
    \includegraphics[width=0.9\linewidth]{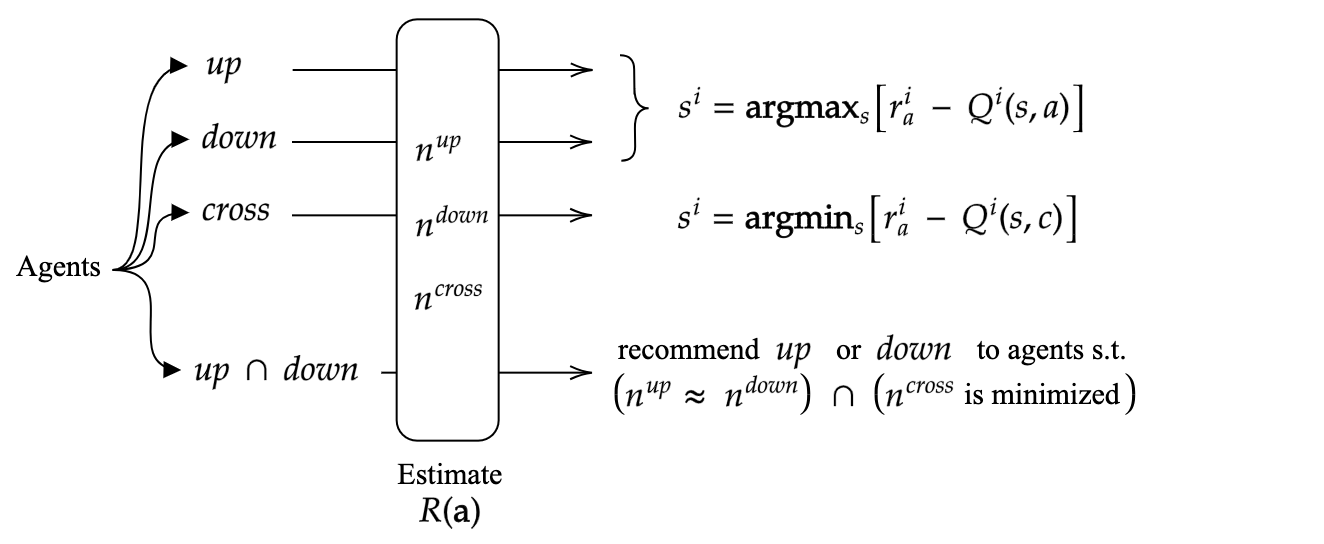}
    \includegraphics[width=0.9\linewidth]{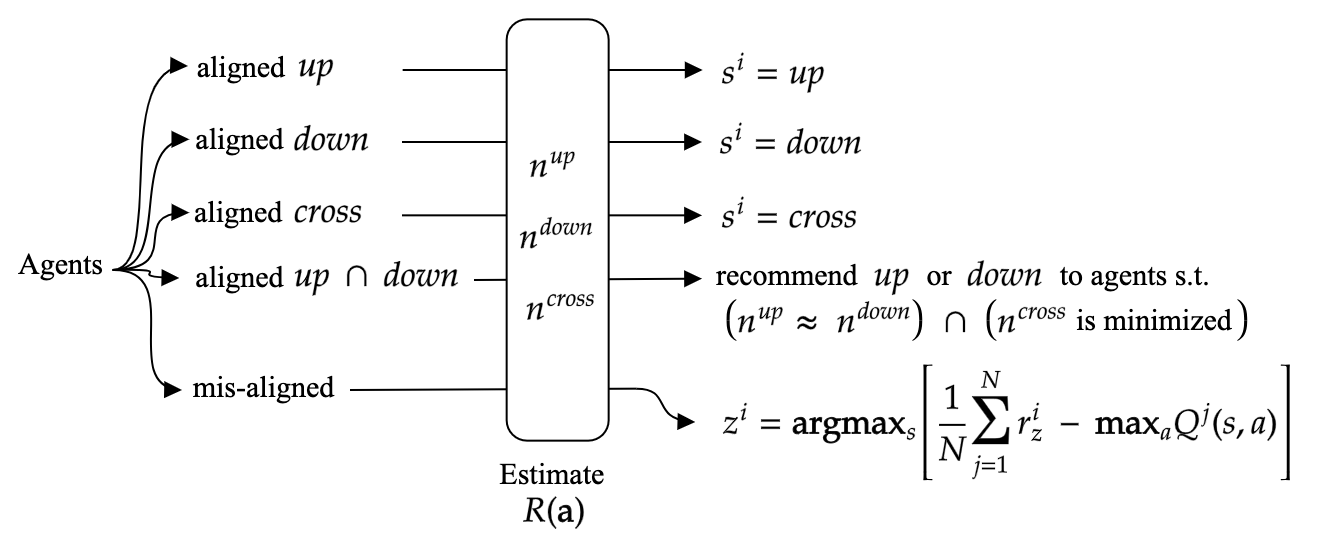}
        \caption{The Heuristic Recommender: Agents are categorized as $\mathit{up}$, $\mathit{down}$, or $\mathit{cross}$ groups according to the $\mathbf{argmax}$ actions in each of their recommendation states. In the case of the $\mathbf{argmax}$ being a tie between both $\mathit{up}$ and $\mathit{down}$, these agents are classified as $up \cap down$. Recommendations are then set such that the, $\mathit{up}$ and $\mathit{down}$ categories' $q$-values would improve most. %while to break any ties between multiple consistent states.
        Similarly, cross agents receive recommendations that worsen their $\mathit{cross}$ $q$-values most. Finally, $up \cap down$ agents are split equally between going up and down, if possible.
        The \textit{Aligned} Heuristic Recommender: agents are categorized as $\mathit{up}$, $\mathit{down}$, or $\mathit{cross}$ groups according to $\mathbf{argmax}=s_{i,t}$, that is that the recommendation $s_{i,t}$ is aligned with the $\mathbf{argmax}$ action. Agents can be classified into multiple groups. For agents that fall exclusively into a single group, they are recommended their aligned recommendation state. %Consequently, their recommendations \textbf{will be aligned}.
        The agents that are both in the $\mathit{up}$ and $\mathit{down}$ groups are given (still aligned) recommendations such that they are split equally between going up and down, if possible.
        Agents with no aligned recommendation states are categorized as misaligned. All misaligned agents receive the same recommendation state, picked such that it is the state whose average belief change over all agents is maximized (the equation in the picture explains it best).
        }
    \label{fig:heuristics}
\end{figure}

In this section we will discuss the case where a RS provides route recommendations to $Q$-learners, so where the number of recommendations $m$ is the number of possible actions for the $Q$-learners $k$. For the initial ($k=2$) and augmented ($k=3$) Braess networks, we explored the potential of using a route recommender to steer the system of agents to achieve the social optimum (least latency). \autoref{fig:heuristics} summarizes the heuristic route recommender for the augmented network.

\subsection{Information Assumptions for the Recommender}

When the $q$-values of the $Q$-learners are known the RS is said to be \textbf{omniscient}. In other words, the RS models the beliefs of agents perfectly, and can determine its precise position in $\sQ\textbf{SPACE}$. With this knowledge, a RS could use a reinforcement learning algorithms (like $Q$-learning) with a polynomial worst-case guarantee $\mathcal{O}(n^c)$ to learn the reward function and transition function, even though the constant $c$ may be prohibitively large. Additionally, if the policies of the $Q$-learners are known, the RS could be seen as having access to a perfect simulator which could be used to create learning samples without needing to interact with the MDP through trajectories. While an unrealistic assumption, it will be useful in the design of our algorithm to imagine that we have knowledge of the $q$-values. If the RS has an inaccurate model of the $q$-values of the $Q$-learners the problem becomes a POMDP, where POMDPs are \textbf{NP}-hard in the worst case \citep{papadimitriou1987complexity, mundhenk2000complexity}. The authors do not know if this particular problem admits an efficiently computable approximation \citep{lee2007makes}, but hypothesize that it does not. For this reason, our algorithm is based on heuristics rather than reinforcement learning. Our experiments assume that: 1) the RS is omniscient, 2) the $Q$-learners have $\epsilon$-greedy policies, 3) the RS models $Q$-learners with greedy policies. Therefore, when the $Q$-learners exploration rate $\epsilon = 0$, the RS correctly models the policies, but when $\epsilon > 0$ the RS has an incorrect model. Thus, $\epsilon$ will represent the likelihood that the RS makes an error in predicting the actions of users. In our experiments we will show what the effects of having an incorrect model of the policies of the $Q$-learners has on the system, while preserving perfect knowledge of the $q$-values.

\subsection{Recommendation Alignment}
\label{sec:rec_aling}

We have thus set the stage to introduce what is meant by recommendation alignment. For a given agent $i$ at iteration $t$, we use the following definitions, assuming there are $|A|$ possible recommendations $s_{i,t}$ corresponding to each possible action, which users treat as state input to their $Q$-function. A recommendation $s_{i,t}$ is \textit{aligned} for user $i$ when $s_{i,t}$ is the action with the highest $q$-value in the state $s_{i,t} \in \mathbb{S}$: $s_{i,t}=\mathbf{argmax}_{a}Q_{i}\left(s_{i,t}, a\right)$. Consequently, a recommendation $s_{i,t}$ is \textit{misaligned} when $s_{i,t}$ is not the action with the highest $q$-value in the state $s_{i,t}$:  $s_{i,t} ~\neq \mathbf{argmax}_{a}Q_{i}\left( s_{i,t},a\right)$.

\paragraph{Recommendation alignment can be a metric for trust}

Given that users are modeled as learners, an aligned recommendation suggests that users have learned to follow the recommendations being given by the RS. This can be taken as an indication that the user \textbf{trusts} the RS. Consequently, a user with a misaligned recommendation does not trust the RS for that particular recommendation. 

\paragraph{The Particular Case of Misaligned Agents}

In the case where an RS knowingly recommends a misaligned recommendation, we define the recommendation as a \textbf{manipulative} recommendation. Furthermore, it is possible for the beliefs of a user to have no aligned recommendations: for the users' $q$-values to lead him \textit{not} to follow any of the recommendations. In such a case, it is not possible to provide the agent with a non-manipulative recommendation. However, it may still be possible to provide a recommendation that may establish alignment during learning.

\paragraph{A Trade-off Between Welfare and Alignment}

The optimal policy $\pi_{RS}^*$ is guaranteed to maximize $G$, and at states $\textbf{s}_t\in\mathcal{S}$ takes action $\mathbf{a}_t = (s_1, \dots, s_N)$. The elements of $\mathbf{a}_t$ may or may not be aligned recommendations for all agents. Therefore, if a $RS$ were to restrict itself from providing misaligned recommendations, it may not be able to recommend the optimal recommendation. As such, for some multi-agent systems, there can exist an inevitable trade-off between recommendation alignment and system welfare.

We have simultaneously conceptualized and demonstrated that the gains in social welfare could come at the loss of recommendation alignment. A misaligned recommendation means that a user has learned not to follow a recommendation, which we interpret as a metric of trust.

\subsection{Solving the Augmented Network with the LDMP}

\paragraph{Mapping latency to Social Welfare}
All subsequent results are displayed in terms of the social welfare, which is a normalized average latency of the network. In the augmented network of the Braess' Paradox a social welfare of 0 corresponds to the worst-case congestion achieved at the Nash Equilibrium where all agents cross for an average latency of $2$. On the other hand, a social welfare of 1 corresponds to the best-case congestion at the socially optimal utilization where half of the agents pick $u$ and half of the agents pick $d$ with an average latency of $1.5$.

\begin{figure}[!h]
    \centering
    \includegraphics[width=0.95\linewidth]{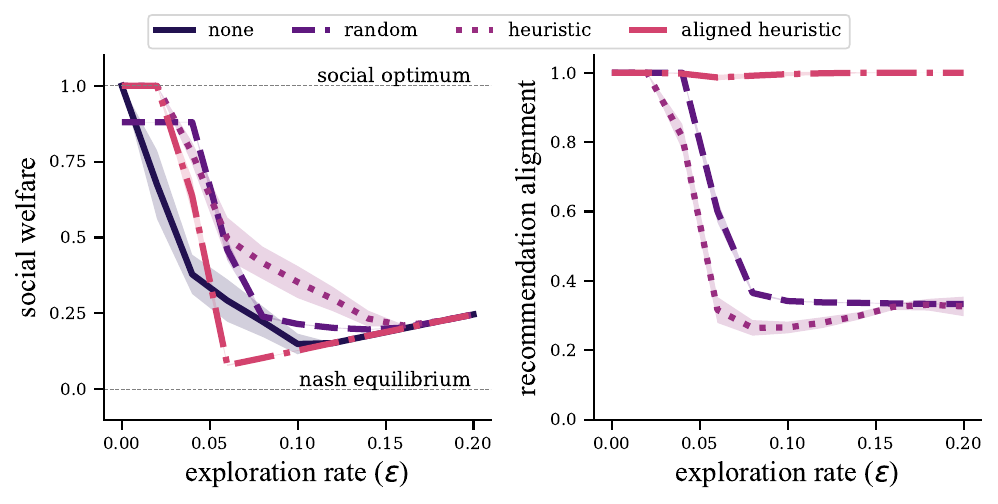}
    \includegraphics[width=0.95\linewidth]{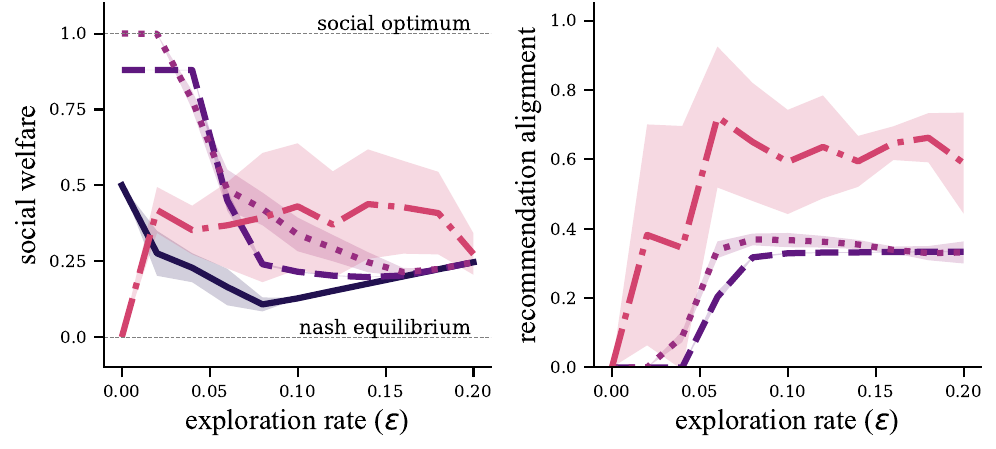}
    \caption{100 $Q$-learners ($\alpha=0.01$, $\gamma=0$) are simulated for 10000 learning steps in the augmented network \autoref{fig:sub2} for varying RS and $\epsilon$ exploration rates. Curves plot averages 40 repetitions of training with error bars as standard deviation. Top row: The systems were initialized with aligned $q$-values (see \autoref{q-values}). The experiments show that the heuristic recommender yields the lowest latencies while using the aligned recommender results in high latencies (left column). However, the aligned recommender also maintains a high alignment with the beliefs of the agents (right column). Bottom row: Initializing misaligned $q$-values (see \autoref{q-values}) leads to a different evolution of the system, which also allows the aligned recommender to have lower latencies than all other recommenders at $\epsilon>0.08$, while maintaining higher alignment.}
    \label{fig:recommender_results}
\end{figure}

\paragraph{Testing different initial beliefs of users}
\autoref{fig:recommender_results} shows the relative performances for these RSs for two cases of initialized $q$-values: aligned $q$-values where the $\mathbf{argmax}$ for each recommendation state corresponds to the recommendation and misaligned $q$-values where the $\mathbf{argmax}$ does not correspond to the recommendation state The $q$-tables are initialized as the following matrix for each agent, aligned and misaligned respectively:

\begin{equation}\label{q-values}
    \begin{bmatrix}
    -1.5 & -2 & -2\\
    -2 & -1.5 & -2\\
    -2 & -2 & -1.5\\
    \end{bmatrix} \text{ and }
    \begin{bmatrix}
    -2 & -1.5 & -2\\
    -2 & -2 & -1.5\\
    -1.5 & -2 & -2\\
    \end{bmatrix} 
\end{equation}
The results for two additional initializations of $q$-values are included in \autoref{appendixC}). Notably, in the misaligned initialization, each agent still has each action $up, down, cross$ as the $\mathbf{argmax}$ of one of its states, but it does not align with the recommendation state.

To compare and contrast different RS approaches we test the following RS which are visible in \autoref{fig:recommender_results}.
(\textbf{none}) The RS produces the constant recommendation $S = (u,d,u,d,\dots,u,d)$. In practice, any constant recommendation is equivalent when all agents have their $q$-values initialized in the same manner. Furthermore, it is identical to the case where no recommendation is provided. This case is to demonstrate that a successful RS must actively, and dynamically change recommendation states for the agents to benefit the social welfare. 
(\textbf{random}) The RS generates a recommendation at each step that is uniformly picked at random from $S$. This case demonstrates that a random RS can have the beneficial effect of slowing down the natural learning dynamics. It also establishes a baseline recommendation alignment score.
(\textbf{heuristic}) The RS attempts to pick recommendations such that as many agents as possible split between up and down, and the remaining agents are recommended actions that would lead their beliefs to change favorably due to the learning dynamics. A detailed picture of this heuristic RS is described in \autoref{fig:heuristics}. This RS achieves higher system welfare at the cost of recommendation alignment. The achieved performance increases the leverage of the manipulative potential of the omniscient RS system, which can be seen by poor recommendation alignment scores. The heuristic recommender results are nearly identical for average latency in the aligned (top row) and misaligned (bottom row) cases of \autoref{fig:recommender_results}. This makes sense, as the heuristic recommender, ``blind'' to recommendation alignment, is not affected by the change in $q$-value initialization of the users.
(\textbf{\textit{aligned} heuristic}) The RS attempts to pick recommendations such that as many agents as possible split between up and down, and the remaining agents are recommended actions that would lead their beliefs to change favorably due to the learning dynamics. A detailed picture of this \textit{aligned} heuristic RS is described in \autoref{fig:heuristics}. This still achieves an improvement in the system welfare while prioritizing recommendation alignment and achieving the highest recommendation alignment scores. When possible, this recommender always recommends aligned recommendations.

The results of \autoref{fig:recommender_results} also lend credibility to the trade-off between welfare and alignment. The best social welfare is achieved by the \textbf{heuristic} RS which did not avoid misaligned recommendations and achieved poor alignment scores. On the other hand, the best alignment scores were achieved by the \textbf{aligned heuristic} RS which achieves lower optimizations of the welfare.

\end{document}